THE 217.5 nm BAND, INFRARED ABSORPTION AND INFRARED EMISSION

FEATURES IN HYDROGENATED AMORPHOUS CARBON NANOPARTICLES

W. W. Duley[1] and Anming Hu[2]

[1]Department of Physics and Astronomy, University of Waterloo, 200 University Ave. West, Waterloo, Ontario N2L 3G1, Canada.  wwduley@uwaterloo.ca

[2] Department of Mechanical and Mechatronics Engineering, Centre for Advanced Material Joining, University of Waterloo, 200 University Ave. West, Waterloo, Ontario N2L 3G1, Canada. a2hu@uwaterloo.ca

ABSTRACT

We report on the preparation of hydrogenated amorphous carbon nano-particles whose spectral characteristics include an absorption band at 217.5 nm with the profile and characteristics of the interstellar 217.5 nm feature. Vibrational spectra of these particles also contain the features commonly observed in absorption and emission from dust in the diffuse interstellar medium. These materials are produced under "slow" deposition conditions by minimizing the flux of incident carbon atoms and by reducing surface mobility. The initial chemistry leads to the formation of carbon chains, together with a limited range of small aromatic ring molecules, and eventually results in carbon nano-particles having an $sp^2/sp^3$ ratio $\approx 0.4$. Spectroscopic analysis of particle composition indicates that naphthalene and naphthalene derivatives are important constituents of this material. We suggest that carbon nano-particles with similar composition are responsible for the appearance of the interstellar 217.5 nm band and outline how these particles can form in situ under diffuse cloud conditions by deposition of carbon on the surface of silicate grains. Spectral data from carbon nano-particles formed under these conditions accurately reproduces IR emission spectra from a number of Galactic sources. We provide the first detailed fits to observational spectra of Type A and B emission sources based entirely on measured spectra of a carbonaceous material that can be produced in the laboratory.

# 1. INTRODUCTION

Recent laboratory simulations of the composition of interstellar carbons have focussed on compositions containing a mixture of sp, $sp^2$ and $sp^3$ hybridized carbon components. These materials can be generated in a variety of ways, including vapor deposition (Wada et al. 2009, Mennella et al. 2003), UV photolysis of solid precursors (Dartois et al. 2005), condensation of particles in a supersonic expansion (Jager et al. 2008, Gadallah et al. 2011), thermal decomposition of precursor gaseous compounds (Steglich et al. 2012) and by heterogeneous condensation of the products of laser ablation (Scott & Duley 1996, Grishko & Duley 2000, Hu et al. 2006). The primary goal of these studies has been to provide accurate spectroscopic simulations of known UV and IR spectral features arising from interstellar carbons and, to a large extent, this goal has been achieved with respect to the reproduction of IR absorption features observed in astronomical spectra (Dartois et al. 2007, Grishko & Duley, 2000, 2002, Duley et al. 2005, Scott & Duley 1996). The results of these laboratory experiments indicates that carbonaceous dust in the diffuse interstellar medium (DISM) and in circumstellar sources has a mixed aliphatic-aromatic composition similar to that occurring in hydrogenated amorphous carbon (HAC or a-C:H) (Sloan et al. 2007, Kwok & Zhang 2011). The relative proportion of aliphatic and aromatic components in this material varies widely, and depends on such factors as thermal history, exposure to radiation and hydrogen content (Angus & Hayman 1988, Robertson & O'Reilly 1987), implying that there is no unique composition for this material under astronomical conditions (Duley et al. 2005, Jones 2012). For example, the composition of the carbonaceous dust seen in the spectrum of CRL 618 (Chiar et al. 1998) is clearly different from that observed toward Galactic center sources (Pendleton & Allamandola 2002) and both of these materials are different from that seen in dense clouds (Brooke et al. 1999). There are, nevertheless, representative compositions that are observed in specific environments. One of these is the composition of HAC which is widely present in the DISM. This material is responsible for the IR absorption bands seen in absorption in the Galactic center (Pendleton & Allamandola 1992) and in the DISM in extragalactic sources (Dartois et al. 2007, Dartois & Munoz-Caro 2007). It must also be chemically related to the component of the DISM that produces the strong extinction feature at 217.5 nm (4.598 $\mu m^{-1}$, 5.71 eV) (Stecher 1965, Savage 1975, Fitzpatrick & Massa 1986) and to the compounds responsible for the IR features at 3.2, 6.2, 7.7, 8.6, & 11.3 $\mu m$ seen in diffuse emission (Sakon et al. 2004).

In this paper, we have determined a set of deposition conditions in the laboratory that replicate, to a large extent, those present in the DISM. HAC deposited in this way contains a component with an absorption band at 217.5 nm whose central wavelength and profile corresponds to those observed along various sight lines in the DISM as determined by Fitzpatrick & Massa (1986). IR absorption spectra of this material also reproduce that seen from the DISM. Analysis of this material indicates that it contains the hydrocarbon functional groups associated with the IR emission features at 3.2, 6.2, 7.7, 8.6, 11.3 $\mu m$ and other wavelengths. We use this data to provide detailed spectroscopic fits to a number of canonical IR emission spectra, including those corresponding to Type A and B sources (van Diedenhoven et al. 2004).

## 2. EXPERIMENTAL

The experimental system was similar to that used in previous ablation/deposition experiments (Scott & Duley 1996, Scott et al. 1997, Grishko & Duley 2000, 2002). Samples were prepared in a vacuum chamber with a base pressure of 5 x $10^{-7}$ Torr. Mass spectra obtained during ablation indicate that H, C, $C_2$ and $C^+$ are the dominant ablation products. A high purity pyrolytic graphite rod was used as the ablation source and this was rotated during deposition to ensure a constant ablation rate. Hydrogen is always present as an impurity in graphite under these conditions and reacts with chemically active components during deposition to saturate any available bonds. Deposition occurred on a fused quartz substrate oriented almost parallel to the target at a distance of 4.5 cm. This could be heated by conduction or cooled to liquid nitrogen temperature. Films were deposited with the substrate held at 600, 300 and 77K. The ablation source was radiation from a XeCl excimer laser (308 nm, 30 ns) operating at a repetition rate of 15 Hz. The incident laser intensity was 3 x $10^8$ W/$cm^2$ corresponding to a fluence of 9 J/pulse $cm^2$ at the target surface. To avoid the condensation of moisture on films deposited at 77K, samples were kept in high vacuum until warmed to room temperature. UV/VIS spectra of these samples were recorded during deposition using an Ocean Optics spectrometer (HR4000 CG) and a deuterium lamp light source.

Raman spectra were obtained using a Renishaw micro-Raman spectrometer with a 50x objective and using an excitation wavelength of 633 nm and a power of up to 3mW. The spectral resolution was 2 $cm^{-1}$ and the laser beam spot size on the target was 5 μm at full laser power. XPS measurements were carried out using a Kratos Ultra photoelectron spectrometer with a monochromatic Al $K_\alpha$ 1486.6 eV X-ray source. The spectrometer was calibrated by Au $4f_{1/2}$ (BE of 84 eV) with respect to the Fermi level. Microstructure characterization involved optical microscopy, scanning electron microscopy (Leo SEM) operating at 15 KV and a commercial atomic force microscope (AFM).

## 3. SAMPLE PROPERTIES

Fig. 1 shows SEM images of films deposited at 300 and at 77 K. It is apparent that temperature has a profound effect on structure, as the film deposited at 77 K consists of nm-sized clusters embedded in a granular matrix, while 300 K films are basically free from granular structure even at high magnification. Films deposited at 600 K have a similar morphology to those deposited at 300 K. SEM images of 77 K films show that these clusters are assembled from a collection of many nm-sized particles. AFM scans of these samples indicate that no substructure exists in films deposited at 600 K at resolutions > 10 nm, while the material deposited at 77 K is found to be assembled from individual grains having sizes $\leq$ 40 nm. These grains combine to form clusters with lateral sizes of $\approx 0.5$ μm, and a height of $\sim 35$ nm.

The nature of carbon bonding in these materials has been determined using x-ray photoelectron spectroscopy (XPS). Measurements indicate that both $sp^2$ and $sp^3$ hybridized bonded carbon is present in all samples, as evidenced by the appearance of peaks at 284.4 and 285.2 eV, respectively. The overall $sp^2/sp^3$ ratio, obtained from the ratio of the strength of these two components, shows that the $sp^3$ content is 20, 72, and 68% in 600, 300, and 77 K films, respectively. This indicates that C=C double bonds dominate only in the high temperature deposit. In all cases, the structure is consistent with the Robertson & O'Reilly (1987) model in which islands of $sp^2$-bonded material are dispersed in a matrix of $sp^3$-bonded carbon.

Raman spectra for the three samples are shown in Fig. 2. Each spectrum is characterized by a combination of G and D peaks in the 1600-1300 cm$^{-1}$ region that can be associated with scattering from sp$^2$ sites. The higher energy G peak arises from the stretching vibration of all C=C bonds (olefinic and aromatic) while the D peak near 1330 cm$^{-1}$ arises only from C=C bonds in aromatic rings (Ferrari & Robertson 2000). An increase in the D/G ratio with sample deposition temperature (Fig. 2) then indicates that the average size of sp$^2$-bonded ring structures is larger in samples deposited at 600 K than at lower temperatures. Spectral components derived from fits to the G and D peaks are summarized in table 1.

The Raman spectrum of the 77 K deposit is particularly simple and is characterized by a G peak at 1551 cm$^{-1}$ and a D peak at 1331cm$^{-1}$. The shift of the G peak to lower energy, together with a D/G ratio ~ 0.6 is consistent with a sp$^3$-bonded carbon concentration of 0.6-0.7 (Ferrari & Robertson 2000). This value also agrees with the sp$^3$ content (0.68) determined from XPS measurements on these samples. The increase in energy of the G band accompanied by a higher D/G ratio in samples prepared at 300 and 600 K occurs as the sp$^2$-bonded component increases relative to the sp$^3$-bonded component. This is not a smooth transition, as evidenced by the high concentration of the sp$^3$ component in samples prepared at 300 K (see Ferrari & Robertson 2000 for a discussion of hysteresis effects in these materials). XPS measurements indicate that the relative concentration of the sp$^3$-bonded component has decreased to ~ 0.2 in samples prepared at 600 K. This result, together with a shift in the G band to ~ 1595 cm$^{-1}$ and a D/G ratio between 1.0 and 1.5, is consistent with the Ferrari & Robertson (2000) correlation. It should be noted at this point that the G and D peaks do not derive from unique chemical structures within these materials, but represent scattering from all the individual chemical components present in the films. Resolution of this structure to gain further insight into the detailed chemical composition of these deposits can be obtained using surface enhanced Raman spectroscopic (SERS) techniques.

As part of the characterization of these deposits, we have measured the optical band gap, $E_g$ in each sample. The optical band gap in amorphous semiconducting films can be obtained from the energy dependence of the absorption coefficient $\alpha$ (Tauc et al. 1966), where $\alpha$ is derived from the transmittance $T = (1-R)^2 e^{-\alpha d}$. $R$ is the reflectance while the film thickness, d, was typically 50-200 nm in our samples. The transmittance was measured at various film thicknesses in order to minimize the effect of reflectance. Absorption data was obtained for samples deposited at different temperatures so that the optical band gap, $E_g$, could be found using standard techniques by extrapolation of the linear part of the following plot to $E = 0$,

$$(\alpha h \nu)^{1/2} = B(E - h\nu)$$

where B is the density of states and hv is photon energy. Films deposited at 600 K have the lowest band gap energy (0.35 eV), while $E_g$ is 1.27 eV and 0.95 eV for films deposited at 300 K and 77K, respectively. All of these materials are therefore semi-transparent at visible wavelengths. From an analysis of the optical band gap energy, it can be concluded that the $sp^2/sp^3$ ratio governs the optical band gap of the films produced in this set of experiments. The slightly lower band gap energy in films deposited at 77K compared to those deposited at 300 K, suggests that there may be enhanced coupling between $sp^2$ clusters through the $sp^3$ bonded matrix in the 77 K material.

Absorption in these materials is determined by the composition of their constituent chemical groups. The net absorption coefficient, $\alpha$, at a given energy will be the product of overlapping absorption from all these components. For materials containing a range of polycyclic aromatic hydrocarbon groups, absorption at low energy is a function of the presence of $sp^2$-bonded components containing many $C_6$ rings (Robertson & O'Reilly 1987) and the optical band gap energy is constrained by the components containing the largest number of rings. This may not be the case in the present samples, particularly those prepared at 77 K, because Raman spectra indicate that a limited number of 1-3 $C_6$ ring molecules are favored under these preparation conditions. From the NIST Chemistry WebBook, maintained by the National Institute for Standards and Technology[1], absorption at higher energies ($h\nu \geq 4$ eV) is primarily due to molecules with a small number of rings and to non-aromatic hydrocarbon chromophores. In this context, the simplest $C_6$ aromatic ring (benzene) has its most intense absorption at ~ 7.1 eV while naphthalene (two rings) has this at ~ 5.6 eV and anthracene (three rings) absorbs strongly at ~ 5.2 eV. Substitution at the periphery of these molecules will tend to reduce these energies to some extent. Small alkane and alkene chains such as propane, $C_3H_8$, and propene, $C_3H_6$, have their absorption bands at ~ 8 and ~7 eV, respectively (Albrecht et al. 2003), while 1,3-butadiene has a strong absorption band at ~ 5.7 eV (Saltiel et al. 2001). It is important to note that polyynes, $C_nH_2$, also absorb strongly in the UV and there is evidence that these groups are present under certain conditions (Hu et al. 2007). For example, $C_8H_2$ produces three intense absorption features near 5.6 eV (Cataldo et al. 2008). Individual deposits will contain a range of $sp^2$ and $sp^3$-bonded carbon components producing absorption at photon energies > $E_g$. The distribution of chemical structures subject to these limitations on the number of carbon atoms will determine the wavelength dependence of absorption in these materials. For example, if one or two ring molecules are favored in a particular solid, then

[1]. At http://webbook.nist.gov/chemistry/.

this will produce a distinctive absorption band at short wavelength. If molecular sizes are distributed over a wider range, then the absorption will be relatively neutral. Searching for a material that has a distinctive absorption band at 217.5 nm (5.7 eV) then involves finding a composition in which one or more molecular groups predominate. This is consistent with the model of Cecchi-Pestellini et al. (2008), who have demonstrated that a fit to the central wavelength and width of the 217.5 nm feature can be obtained by combining spectra from a relatively small number of individual PAH components.

The molecular composition of these materials can be effectively probed using SERS techniques. The SERS method utilizes the presence of Ag nano-particles to enhance scattering from highly localized regions at or near the surface of deposited samples (Moskovits 1985). SERS can sample individual molecules and reveals details of both Raman and IR spectra normally obscured by heterogeneous broadening effects (Moskovits, DiLella, & Maynard 1988). It has been shown (Matejka, et al. 1996) that the SERS interaction does not have a significant effect on the energy of the enhanced transitions, and that both IR and Raman features occur in SERS spectra. We have previously reported on the use of SERS to investigate IR spectra for comparison with emission spectra of "UIR" sources (Hu et al. 2006, Hu & Duley 2007, 2008a,b). Here, we use the same techniques to see if we can obtain more information on the type of chemical structures in the present deposits for correlation with measured UV spectra of these materials.

Fig. 3 shows representative SERS spectra in the 1700-1000 cm$^{-1}$ region for samples deposited at 77, 300 and 600 K. Spectral lines associated with specific molecular groups appear in these scans and the energies of these lines (Table 2) can be used to provide a better definition of the chemical components in the deposits.    Spectra in the region of the G band show that non-aromatic C=C groups are important in material deposited at 600 K. These can be associated with olefinic bridging groups between islands of $sp^2$ bonded rings. Deposition at 600 K results in dehydrogenation and tends to favor the formation of chains containing C=C groups rather than the C−C bonds in $sp^3$ bonded chains in lower temperature deposits. The G band itself is found to contain a number of components. The most intense feature occurs at 1591, 1603 and 1601 cm$^{-1}$ in the 77, 300 and 600 K deposits, respectively. This band can be associated with the C=C stretching bond in rings, although it is also possible that it could be formed by C=C bonds in chains. The slightly lower energy of this feature in the 77 K deposit may signal the presence of somewhat smaller rings in this material than in samples formed at higher temperatures.

The spectral features listed in Table 2b are assigned to a range of substituted small aromatic ring compounds based on extensive compilations of Raman spectra (Socrates 2001), together with the appearance of a peak near 1000 cm$^{-1}$ which is highly characteristic of substituted benzene rings. This feature is weak in the spectra shown in Fig. 3, but is often the strongest band in other samples. Standard compilations indicate that the spectral feature at 1377 cm$^{-1}$ is typical of substituted naphthalenes, but many other ring molecules also have a spectral feature at this energy. Other bands show clear evidence for the presence of structures consisting of two benzene rings joined by a variety of bridging groups. Notable components may include: diphenyl acetylene (DPA), diphenylethylene (DPE), diphenylbutadiene (DPB), diphenylhexadiene (DPH), and biphenyl (B). Larger ring structures are likely also present with and without substituents. Some possibilities include binapthyl (BN), o-terphenyl (OT), phenanthrene (PN) and triphenylene (TP). CH$_n$ groups are apparent in the low temperature deposits and may be in the form of side-chains on the small rings or as components of a hydrocarbon matrix. Time of flight (TOF) mass spectrometric studies of the decomposition of these materials (Scott et al. 1997) has shown that small carbon molecules containing < 20 carbon atoms including alkane and olefinic chains are abundant in these deposits.

Additional information on the molecular constituents of these samples can be obtained from SERS spectra at energies < 1000 cm$^{-1}$.  A variety of ring deformation modes and modes associated with CH$_n$ groups are found in this range (Socrates 2001). These can be used to further characterize ring substituents and peripheral groups within the chemical building blocks in materials deposited at different temperatures.   Fig. 4 shows spectra in this energy range of samples deposited at 77 and 600 K and the energies of specific features appearing in these spectra are listed in Table 4, together with possible assignments.  These assignments are obviously not unique given the complexity of organic chemistry, but correspond to simple functional groups as inferred from the analysis of spectral features at higher energy (see Table 2). The spectra have features that lie in the spectral range corresponding to bending vibrations of PAH rings with solo, duo, trio and quartet CH.

The nature of chains attached to rings or chains acting as bridges between rings in these materials can be obtained from Raman spectra in the energy range 1900-2300 cm$^{-1}$.  Raman spectra of the 77 K sample (Fig. 5), shows features at 1977, 2087, 2113, 2130 and 2233 cm$^{-1}$, while the 300 K sample has bands at 1941, 1992, 2002 and 2137 cm$^{-1}$ (weak).  The sample deposited at 600 K has features at 1995 and 2025 cm$^{-1}$.  Previous studies (Ravagan et al. 2002, D'Urso et al. 2006, Lucotti et al. 2006, Szepanski et al. 1997) have shown that spectral features in this energy range can be assigned to CC stretching modes in polyyne and cumulene chains. From these correlations, the Raman features at 2087, 2113, 2130 and 2233 cm$^{-1}$ in the 77 K sample and that at 2137 cm$^{-1}$ in the 300 K sample are assigned to polyyne chains (-C≡C-)$_n$, while the bands at lower energies in all samples would appear to be due to cumulenic chains $(C=C)_n$. It is uncertain if these chains are separate entities within the solids or whether they are attached at one (or both ends) to other chemical groups/rings. Fig. 5 shows that the energy of the most intense band arising from cumulenes increases from 1977 to 2025 $cm^{-1}$ as the deposition temperature increases from 77 to 600 K. This indicates that the chain length decreases in higher temperature samples as the number of aromatic ring structures increases. These spectra, with a strong band at 1975 $cm^{-1}$, are consistent with SERS spectra of carbon chains prepared by electric arc discharge in methanol (Lucotti et al. 2006). Analysis of Raman scattering by carbon chains in carbonized fluoro-polymers (Kastner et al. 1995) has shown that the energy, $\nu(N)$, of the CC stretching mode in polyyne chains occurs at $\sim \nu(N) = 1750 + (3980/N)$ $cm^{-1}$, where N is the number of C atoms. If this correlation also holds in the present samples, then we find that $N \approx$ 12, 11, 10 and 8 for the peaks at 2087, 2113, 2130 and 2233 $cm^{-1}$ in the 77 K samples, and $\approx 10$ in the 300 K sample. However, this correlation is strongly affected by the termination at the end of the chains. For example, from the SDBS WebBook[2], the energy of the $C\equiv C$ band in diphenylacetylene (where N = 2) is found to be 2225 $cm^{-1}$, suggesting that the band observed at 2233 $cm^{-1}$ in the 77 K deposit probably arises from diphenylacetylene, rather than from a separate polyyne chain.

[2]. SDBSWeb : http://riodb01.ibase.aist.go.jp/sdbs/ (National Institute of Advanced Industrial Science and Technology,

Spectra of samples deposited at 77 K also have several weak features between 2600 & 2800 cm$^{-1}$ which have been shown to be characteristic of CH$_n$ groups (Lawson et al. 1995). These are overtone and combination bands involving deformation modes of CH$_3$ and CH$_2$ groups between 1300 and 1400 cm$^{-1}$. These weak features are seen in the spectrum of many different hydrocarbons, but the energy of individual features can be closely related to bonding environment. In our spectra, (Fig. 6) bands are seen at 2780, 2722, 2700 and 2615 cm$^{-1}$. Of these, the band at 2722 cm$^{-1}$ can be identified with 1,3 dimethyl-naphthalene and 2,6 dimethyl-naphthalene and arises from methyl groups substituted at various locations on the double ring. The feature at 2780 cm$^{-1}$ may be associated with a branched chain such as 2,2 dimethyl-hexane containing tertiary carbon atoms or a substituted ring structure such as tert-butyl benzene. The other bands at 2700 and 2615 cm$^{-1}$ would arise from branched alkane structures containing a variety of CH$_n$ groups.

Fig. 6 also covers the spectral region corresponding to the primary IR and Raman bands of CH, CH$_2$ and CH$_3$ groups. A notable aspect of these spectra is the absence of the asymmetric stretching band of CH$_3$ which is typically seen at 2960 ± 20 cm$^{-1}$. The absence of CH$_3$ features in our spectrum may arise from a vagary of the SERS process itself, in which signals from certain functional groups are suppressed, as CH$_3$ bands are observed in IR absorption spectra of these samples (Duley et al. 1998). The strongest components in all three spectra in Fig. 6 occur at 2925 ± 3 cm$^{-1}$ and 2860 cm$^{-1}$. These correspond to the asymmetric and anti-symmetric stretching modes, respectively, of methylene (CH$_2$) groups. The spectrum of the 77 K deposit also shows an additional band at 2882 cm$^{-1}$ with a FWHM of 40 cm$^{-1}$. This band can be associated with a Fermi resonance involving an overtone of the CH$_3$ bending vibration near 1440 cm$^{-1}$ and the symmetric stretching band of CH$_3$, but it is also possible that it arises from tertiary CH groups. Tertiary CH groups appear at the surface of diamond when a surface atom is hydrogenated, and the observation of a 2880 cm$^{-1}$ band in dense interstellar clouds led to the suggestion that diamond particles are present in these objects (Allamandola et al. 1992). We will discuss the possibility that nano-diamond is a component of these samples in section 5.

Fig. 6 also shows a weak feature at 3060 ± 5 cm$^{-1}$, which can be assigned to the CH stretching mode in aromatic CH groups. Many PAH molecules have a CH stretching band in this energy range, making it surprizing that the weak feature in Fig. 6 is so well defined, since the presence of a multitude of PAH structures would be expected to produce a broader feature. This is further evidence that samples contain a limited range of sp$^2$ bonded structures and that only a few of these are dominant. A variety of substituted naphthalenes including binapthyl have a Raman feature at 3060-3070 cm$^{-1}$.

The picture that emerges from these spectra is that these samples consist of a collection of relatively small aromatic ring structures in conjunction with a framework of acetylenic, aliphatic and olefinic chains (see Jones et al. 1990). These chains contain several carbon atoms, although it is possible that larger chains may also be present. The UV absorption spectrum of these materials will reflect this composition. Since Raman data indicates that these samples are formed from a limited number of relatively small molecular building blocks, the strongest absorption features in the spectrum will arise from the most abundant of these molecules.

# 4. UV SPECTRA

Fig. 7 gives a comparison between the UV absorption spectrum of films deposited at 77, 120, 300 and 600 K. It is apparent that a noticeable effect of deposition at higher temperature is to increase absorption at longer wavelengths relative to that at $\lambda < 250$ nm. Spectral features at ~230, 270, 280, 310 and 360 nm in the 600 K deposit, and that at ~ 240 nm in the 300 K deposit are only weakly present in the 77 K sample. It seems that deposition conditions are such that a similar range of molecular constituents can form at 77 K and higher temperatures, but that the final distribution of components is affected by the overall temperature. Both the 77 and 600 K deposits have a well defined absorption band at ~ 230 nm (5.3 eV). A notable feature of the spectrum of the sample deposited at 77 K is the appearance of an additional absorption band at ~ 220 nm. This feature is characteristic of samples deposited at lower temperature, but it is also found in some samples deposited at 300 K.

Most samples deposited at < 150 K exhibit a 217.5 nm absorption band in the form of a distinct feature relative to the broader absorption contributed by other absorbing groups produced under these conditions. This suggests that the concentration of the carrier of the 217.5 nm band is enhanced by deposition at low temperature, but it is apparent that the carrier can also be formed at higher temperature, since a weak 217.5 nm band also occurs in samples deposited at $\geq 300$ K. Evidently, the chemistry leads to the formation of a few dominant chemical constituents including the chromophore that is the source of the 217.5 nm feature. Slight changes in the central wavelength of this feature may reflect different substitution and bonding on the periphery of this molecule. Beegle et al. (2007) have found that a $\pi \rightarrow \pi^*$ transition appears near 217.5 nm in UV spectra of naphthalene aggregates and chemical derivatives.

The observation of discrete absorption bands in these samples over this wavelength range, rather than continuous absorption, indicates that they contain a limited distribution of relatively small molecular components. We find that the relative strength of individual band components in these UV spectra is variable reflecting the detailed kinetics of the deposition process as well as the substrate temperature and gas composition. A summary of peak absorption wavelengths, $\lambda_m$, for some possible molecular components is given in Table 4. These specific components are indicated by the appearance of spectral features in the SERS scans (Figs. 3-6).

More information about the composition of the chemical components that need to be present to produce a 217.5 nm absorption band in these samples can be obtained by comparing Raman spectra of samples deposited at 77 and 600 K. Inspection of Fig. 3 shows that features at 1578, 1467, 1432, 1287, 1275, 1264 and 1216 cm$^{-1}$ are enhanced in the 77 K deposit relative to that at 600 K. Although there is some ambiguity due to the complexity of molecular spectra in this region, all of these features can be associated with naphthalene and its singly and doubly substituted derivatives including methyl- and dimethyl-naphthalene and other hydrocarbon derivatives such as ethyl-naphthalene. This suggests that these components are more abundant in the 77 K deposit than in samples prepared at higher temperatures and implies that there may be a connection between the presence of these relatively simple compounds and the appearance of a 217.5 nm absorption band. In this context, the CH$_3$ overtone/combination bands seen at 2780, 2722 and 2700 cm$^{-1}$ in the spectrum of samples deposited at 77 K are not present in samples deposited at 300 or 600 K indicating that there is a reduction in the concentration of substituted naphthalene compounds in samples formed at higher temperatures.

# 5. COMPOSITION

Spectroscopic analysis of these deposits at wavelengths extending from the IR to the UV under a variety of different excitation conditions shows that they contain a limited range of compositions based on hydrocarbon chains and small aromatic ring molecules. The concentration of $sp^2$-bonded carbon in these solids has also been determined from XPS measurements and is $\sim 0.3$, $\sim 0.3$ and $\sim 0.8$ for samples deposited at 77, 300 and 600 K, respectively. This information will now be used to further define the chemical structure of these materials.

To put these numbers in context, the $sp^2$ concentration in aromatic molecules such as benzene ($C_6H_6$), naphthalene ($C_{10}H_8$), phenanthrene ($C_{14}H_{10}$), etc. is 1.0. This would become 0.75, 0.83 and 0.875 for dimethyl substituted benzene, naphthalene and phenanthrene, respectively. Reducing the $sp^2$ concentration to 0.3 as observed in the 77 and 300 K deposits requires an additional $sp^3$-bonded component, either in the form of hydrocarbon groups bonded to these rings, separate chains or in the form of diamond-like carbon. For all these possibilities, the requisite number of $CH_{2,3}$ groups per PAH would be 14, 23, and 33 for benzene, naphthalene and phenanthrene, respectively. The presence of heavily substituted PAH building blocks in the present samples is not supported by spectral data (see tables 2 & 3) implying that the samples likely have a mixture of partially substituted PAH structures together with separate alkane chains, or that some fraction of this $sp^3$ component is in the form of nano-diamond inclusions.

If hydrocarbon chains are produced in these samples and form a matrix separating substituted PAH molecules, then the maximum number of carbons atoms per chain is determined by the measured $sp^2$ content. Thus, for the above PAH components, the largest hydrocarbon chain would correspond to isomers of dodecane ($C_{12}H_{26}$), docosane ($C_{22}H_{44}$) and dotriacontane ($C_{32}H_{66}$), respectively. Of course, the $sp^3$ chain component could be broken up into several smaller hydrocarbon isomers. Table 3 lists a number of spectral features that are likely attributable to $CH_n$ groups, but it is uncertain if these groups are contained in separate hydrocarbon chains or are substituents on rings. IR absorption spectra of samples produced in the same manner as the present deposits show that the concentration of methyl ($CH_3$) groups is enhanced when deposition occurs at 77 K compared to deposition at 300 K (Duley et al. 1998). As the methyl group must always be at a terminal location, this indicates that heavily branched hydrocarbon and short alkane chains dominate in material formed at 77 K. An enhancement in the relative concentration of methyl and dimethyl-substituted PAH components is also consistent with this result. A reduction in the concentration of methyl groups in 300 K deposits implies that these rings are becoming larger as the samples become less fully hydrogenated. Since these deposits grow by accretion of atoms and molecules ablated from the graphite target, the final composition reflects the higher mobility of atoms on the surface at 300 and 600 K than at 77K. Accreted carbon atoms are then able to migrate farther enabling the growth of larger $sp^2$ bonded structures through carbon insertion reactions which are facilitated at higher temperature. Mobility is limited at 77 K so that the growth of aromatic ring structures will be truncated and only smaller PAH components are formed. When this occurs, the resulting composition is enhanced in the $sp^3$ bonded carbon component.

Extensive measurements of IR spectra of samples with mixed $sp^2/sp^3$ composition (Pino et al. 2008) have shown that there is a good correlation between the energy of the C=C stretching band near 1600 cm$^{-1}$ and the $[CH]/[CH_n]$ ratio which reflects the relative concentration of aliphatic groups. This ratio is approximately equivalent to the $sp^2/sp^3$ ratio measured from XPS spectra, although it does not capture those carbon atoms that are not bonded to hydrogen. If the ratio defined by Pino et al. 2008 reflects the XPS data obtained for our materials, then this predicts $[CH]/[CH_n] = 0.5\text{-}6$ in the 77 K samples and $[CH]/[CH_n] \sim 7$ in those prepared at 600 K. The corresponding $sp^2/sp^3$ ratios from XPS data are $\sim 0.4$ and $\sim 4$, respectively implying that the 77 K samples probably contain a higher concentration of $sp^3$-bonded carbon than those of Pino et al. (2008).

It is also well known that extended regions of $sp^3$-bonded carbon, in the form of a diamond-like phase, is a notable characteristic of amorphous carbon (a-C) films produced by laser ablation and chemical vapor deposition (Malshe et al. 1990, Pappas et al. 1992). The diamond-like phase yields a distinct peak in the Raman spectrum near 1336 cm$^{-1}$ in macroscopic samples. This peak shifts to $\sim 1326$ cm$^{-1}$ in $\sim 5$ nm nano-diamond particles (Prawer et al. 2000). A feature at this energy is seen in the 77 and 300 K spectra (table 2) indicating that these samples may contain nano-diamond inclusions. If this is correct, then part of the "missing" $sp^3$ component in the 77 and 300 K deposits could reside in nano-diamond. If we assume that the $sp^2$ building blocks in these materials are based on naphthalene with one or two peripheral substituents, and that other $sp^3$ bonded components are also present, then a $sp^2$ concentration $\sim 0.3$ would correspond on average to $\leq 20$ $sp^3$-bonded diamond-like carbons per naphthalene unit.

The present measurements show that the $sp^2$ concentration rises to $\sim 0.8$ in samples deposited at 600 K. This would be accompanied by an increase in the average number of rings in the $sp^2$ building blocks and an attendant decrease in the optical band gap in these materials. We note that the energy of the spectral feature in the 77 K deposit tentatively identified with a nano-diamond component (1329 cm$^{-1}$, Table 2) apparently shifts to 1322 cm$^{-1}$ in the spectrum of the 600 K sample. A shift of this magnitude is inconsistent with the data of Ager et al. (1991) since the width of the 1322 cm$^{-1}$ feature in our spectrum (15 cm$^{-1}$) is much less than that observed for nano-diamond particles of the requisite size. In view of this finding, we conclude that there is no clear spectroscopic evidence for diamond nanoparticles in the 600 K deposits.

The concentration of the molecular component giving rise to the 217.5 nm absorption feature in these samples can be obtained from the optical depth, $\tau = \sigma\, n_a\, L = \alpha\, L$ where $\sigma$ is the absorption cross section per absorber, $n_a$ is the number of absorbers cm$^{-3}$, $\alpha$ is the absorption coefficient cm$^{-1}$, and L is the sample thickness. From the absorption data, we obtain $\alpha_{217.5} = 2.2 \times 10^4$ cm$^{-1}$ by assuming that the 217.5 nm absorbers only contribute the observed absorption in excess of the background. With $\sigma = 2 \times 10^{-16}$ cm$^2$ as representative of the absorption cross section for a strong $\pi \rightarrow \pi^*$ transition in a molecule such as naphthalene, $n_a = 1.05 \times 10^{20}$ cm$^{-3}$. Since there are 10 carbon atoms per absorber (we do not include any carbon atoms in side chains), $n_c = 1.05 \times 10^{21}$ cm$^{-3}$ is the density of carbon atoms in the 217.5 nm absorber in these samples. This constitutes $\sim 2$ % of the total density of carbon atoms in these materials. As the sp$^2$ content at 77 K is found to be $\sim 0.3$ from XPS measurements, we find that the 217.5 nm absorbers in the deposits comprise $\sim 7\%$ of the total number of sp$^2$ bonded carbons. If, in the above calculation, we assume that *all* of the absorption at 217.5 nm arises from these absorbers, then $n_c = 3.7 \times 10^{21}$ cm$^{-3}$. This would represent $\sim 7.4\%$ of the total carbon density and $\sim 24.7\%$ of sp$^2$ bonded carbons.

## 6. COMPARISON WITH OBSERVATION

An absorption band at 217.5 nm, commonly associated with a ubiquitous carbonaceous material, is widely observed in the diffuse interstellar medium (DISM). The profile of this feature is primarily Lorentzian (Savage 1975, Seaton 1979) with a peak wavelength that varies slightly (≤ 1%) from one line of sight to another and a width that has been found to change by up to ~ 25%. (Fitzpatrick & Massa 1986). The optical depth, $\tau$ (217.5) = 1.17 $A_v$, where $A_v$ is visual extinction, implying that the 217.5 nm absorber must contain a significant fraction of available carbon. The DISM is also characterized by the appearance of IR absorption bands at ~ 3.4, 6.86 and 7.27 $\mu$m ( ~ 2940, 1460 and 1376 cm$^{-1}$, respectively) arising from the stretching and bending modes of methylene and methyl ($CH_{2,3}$) groups. Another absorption band associated with the stretching mode in C=C bonds in olefinic or aromatic hydrocarbons appears near 6.2 $\mu$m (1610 cm$^{-1}$), but there is no absorption detected at the wavelength of the well-known 3.29 $\mu$m (3040 cm$^{-1}$) aromatic CH IR emission feature. As a result, Dartois et al. (2005, 2007) conclude that the material producing these IR bands in the DISM is highly aliphatic and contains little aromatic material. Their abundance limit, based on the absence of absorption due to aromatic CH at 3.29 $\mu$m, is N(H)$_{arom}$ / N(H)$_{aliphatic}$ < 0.08, where N(H) is column density. It is estimated that the hydrocarbon dust in the DISM contains ~ 15% of available carbon.

This information points to a connection between the material responsible for the 217.5 nm feature and that giving rise to IR absorption in the DISM. In particular, it suggests that the appearance of a band at 217.5 nm may be related to the presence of hydrogenated amorphous carbon, rather than to dehydrogenated carbons such as graphite. Criteria for the assessment of this hypothesis include: i) the proposed material must produce a 217.5 nm absorption band with a central wavelength, profile and amplitude consistent with that of the interstellar feature, ii) the strength of this band must be compatible with the available carbon abundance in the DISM, iii) its IR absorption spectrum must accurately reproduce that of DISM dust and be consistent with the derived abundance of methyl and methylene groups, iv) there must be some indication of how this material is involved in the production of emission of the discrete IR features at 3.4, 6.2, 7.7μm, etc. in the DISM and other sources.

A 217.5 nm absorption feature is present in many of the UV spectra and is undoubtedly related to one or more molecular components. Fig. 8a shows the profile of the 217.5 nm band as it appears in a sample deposited at 77 K and measured at 300 K. The Lorentzian fit corresponds to $\lambda_0^{-1} = 4.57$ μm$^{-1}$ with a FWHM $\gamma = 0.8$ μm$^{-1}$ (5.67 eV and 0.99 eV respectively). These parameters fall within the range of correlations reported by Fitzpatrick and Massa (1986) for the 217.5 nm band in astronomical sources. The spectrum of this sample has an additional weak feature at $3.8 - 4.0$ μm$^{-1}$ (4.71− 4.96 eV), indicating that the material generated in the laboratory has a somewhat different composition from that present in the interstellar medium. The corresponding absorption in another, hydrogen-rich sample is shown, together with its Lorentzian fit in Fig. 8b. In this case, the fit parameters are $\lambda_0^{-1} = 4.57$ μm$^{-1}$ and $\gamma = 1.28$ μm$^{-1}$ (5.67 eV and 1.59 eV respectively) which agree with those derived from a fit to the 217.5 nm profile in the spectrum of ζ Oph (Fitzpatrick & Massa 1986). Fig. 8c shows the absorption spectrum of another laboratory sample where $\lambda_0^{-1} = 4.63$ μm$^{-1}$ and $\gamma = 1.0$ μm$^{-1}$ (5.74 eV and 1.24 eV respectively). These parameters lie at the short wavelength limit of the Fitzpatrick & Massa (1986) fits. This spectrum also has an additional band at 285.7 nm with $\lambda_0^{-1} = 3.50$ μm$^{-1}$ and $\gamma = 1.1$ μm$^{-1}$ (4.34 eV and 1.36 eV respectively).

As discussed, the carrier of the 217.5 nm absorption band in the laboratory samples requires 2-7% of the available carbon and 7-25% of $sp^2$-bonded carbon. If we assume that the interstellar feature at 217.5 nm is produced by the same type of molecular carrier, then we can determine the analogous fraction of available carbon in the DISM needed to give rise to this feature in interstellar extinction. Observations indicate that $\tau$ (217.5) = 1.17 $A_v$, while the column density of hydrogen nuclei N(H) = $1.9 \times 10^{21} A_v$ cm$^{-2}$ in the DISM (Whittet 2003). If N(C)/N(H) = $3.7 \times 10^{-4}$, the fractional carbon abundance in dust is N($C_{dust}$) / N(H) = $3.7 \times 10^{-4} \beta$ where $\beta \approx 0.3$ is determined by the observed interstellar depletion of carbon. Using $\sigma = 2 \times 10^{-16}$ cm$^2$ as before, and taking 10 carbon atoms per absorber, $\tau$ (217.5) = $1.66 \times 10^{-18}$ N(C) = $1.66 \times 10^{-18}$ [N($C_{dust}$) / $\beta$] = $\sigma$ N(abs), where N(abs) = $n_a$ L is the column density of the carrier of the 217.5 nm band in the DISM. Then, after converting to densities, $n_a$ / $n_c$ (dust) = $8.3 \times 10^{-3}$/ $\beta$ and $n_c$(abs) / $n_c$ (dust) = 0.083 / $\beta$, where $n_c$(abs) = $10 n_a$ is the number of carbon atoms cm$^{-3}$ in the DISM dedicated to the 217.5 nm absorber. With $\beta$ = 0.3, one gets $n_c$ (abs) / $n_c$ (dust) = 0.28 corresponding to 28 % of available carbon contained in the 217.5 nm absorber in DISM material. Since this represents the $sp^2$ bonded carbon component, the fraction of $sp^2$ carbon in DISM dust is ~ 0.3.

Finding that 28% of the carbon in DISM dust is required to yield the observed strength of the 217.5 nm feature, while the carrier of this band in laboratory samples contains only 2-7% of the available carbon has strong implications for the composition of interstellar carbon dust. It suggests that carbon dust in the DISM has the same $sp^2$ content (~ 0.3) as the material produced at 77 and 300 K in the laboratory, but that the $sp^2$ component is undiluted by other, presumably larger, PAH groups. As we have seen, laboratory samples contain the $sp^2$ component giving rise to the 217.5 nm feature, but this is only one of a number of alternative $sp^2$-bonded structures. These other components produce absorption at longer wavelengths (see Fig. 7) and do not appear to be present in DISM dust. Since the 217.5 nm band is seen in both laboratory and DISM dust, it appears that DISM carbon dust has a similar $sp^2$ / $sp^3$ ratio ($\sim 0.43$) as that generated in the laboratory, but that the composition of DISM material has been concentrated in some way such that only a very limited variety of $sp^2$ bonded structures are present. This finding is in agreement with a model in which carbon dust can form by accretion on a surface at low temperature where the mobility of carbon atoms is inhibited. It also implies that carbonaceous dust, with a composition similar to that of the 77 K deposits (see Jones 2012), forms in the DISM on a timescale that is comparable to that for dust destruction in shocks ($\leq 10^8$ years) (Jones & Nuth 2011). A real-time formation process, operating under DISM conditions, in which newly formed carbonaceous structures are desorbed from silicate dust or other nuclei may then be another reason why ring formation is truncated before larger structures can appear.

The laboratory data then indicate that carbonaceous dust in circumstellar envelopes of evolved stars may contain a wider variety of PAH and other hydrocarbon compounds than appear in DISM dust. This would be the result of formation at higher density and temperature as both of these parameters facilitate the production of larger molecular groups. In this material, the $sp^2$ bonded carbon component would be distributed over a range of olefinic and aromatic compounds, and would not be concentrated in smaller rings as seems to be the case for dust in the DISM. As a consequence, the 217.5 nm feature, while likely still present, would be obscured by the $\pi \rightarrow \pi^*$ transitions of the other $sp^2$ bonded constituents. The overall distribution between $sp^2$ and $sp^3$ bonded carbons is expected to depend on the processing temperature, with the $sp^3$ hydrocarbon content decreasing significantly in high temperature material. IR spectra will also be different, both in absorption and in emission. In particular, our data suggests that the $CH_3$ band near 3.38 μm (2960 cm$^{-1}$) should be weaker relative to the $CH_2$ band at 3.42 μm (2920 cm$^{-}$ [1]) when observed in emission rather than absorption, as this implies a higher dust temperature. These characteristics of the 3.4 μm emission and absorption band together with a comparison between laboratory and observational data in a variety of sources have been discussed previously (Duley et al 2005).

An extensive compilation of the spectral properties of individual PAH molecules combining theoretical and laboratory (argon and neon matrix) data is now available (Bauschlicher et al. 2010) and has been used to simulate IR emission spectra of PAHs in astronomical sources. The large number of variables associated with this database ensures that a fit can be obtained to any specific astronomical spectrum, but the resulting simulation provides little detailed chemical or physical information. This method also presumes that the emitters are individual gas-phase molecules; an hypothesis that has not been validated by observation (Gredel et al. 2011, Searles et al. 2011, Galazutdinov et al. 2011, Pilleri et al. 2009). This model is predicated on the requirement that the IR emission features are excited by the absorption of individual photons from the ambient radiation field, but recent studies have shown that this is probably not necessary (Duley & Williams 2011). The present situation with regard to simulation of IR emission spectra must then be considered to be unsatisfactory, as one should be able to prepare samples in the laboratory whose properties correspond to those detected in space, rather than having to rely on multi-parameter fits with theoretical spectra.

The challenge then is to produce real materials whose spectral properties reproduce those of the dust as observed in emission in circumstellar and other astronomical sources. We have shown elsewhere that many of the discrete IR emission features observed in astrophysical sources do appear in laboratory spectra of these samples (Hu et al. 2006, Hu & Duley 2007, 2008a,b). The primary IR emission bands, including those at 3.2, 6.2, 7.7, 8.6, 11.3, 12, 12.8, 13.5 & 14.2 μm in Galactic and extra-galactic sources, are seen in these spectra, and the width of individual spectral features is generally 5-30 cm$^{-1}$, although broader bands are also found in some of these deposits. In a given laboratory sample, slight differences in deposition conditions can result in significant changes in the overall spectrum due to different concentrations of specific molecular groups. Spectra of the current materials contain many individual lines (Fig. 3-6 and Tables 2 & 3) and are representative of those observed in most of our samples. However, it is important to realize that these materials do not have the same overall composition as interstellar carbonaceous dust, although they contain many of the same constituent groups.

With this caveat, Fig. 9 shows a fit to the mid IR emission spectrum of the protoplanetary nebula (PPN) IRAS 01005+7910 (Zhang & Kwok 2011). This fit was obtained using a number of the spectral components listed in Table 2. The widths of these lines were taken to be representative of measured values. The inclusion of lines from samples prepared at different temperature may indicate that this object contains a mixture of dust subjected to a variety of processing conditions, but could just as well reflect difficulties in extracting individual spectral lines from complex laboratory spectra. While not perfect, we believe that this is the first true simulation of an astronomical IR emission spectrum of this kind using data from real materials that can be made in the laboratory. We are assembling a database from measured spectra of many samples that will enable a further comparison with astronomical spectra. As an example, Fig. 10 shows fits to generic Type A and Type B emission spectra, as represented by spectra of IRAS 233133+6050 and HD 44179, respectively, (van Diedenhoven et al. 2004) using our laboratory data. Wavelengths and widths of the corresponding components identified in these fits are summarized in Table 6. It is apparent that Type A and B spectra are produced by dust having similar composition, but that there are some chemical differences between the two types of material. Both of these materials also differ in composition from that appearing in IRAS 01005+7910 (Table 5).

The good agreement between UV and IR spectra of these laboratory samples created by laser ablation and those observed from astronomical objects raises the question of why these materials are such good replicas of those existing in space. To answer this question, we have to look more closely at deposition conditions during formation of the laboratory samples. Samples are generated by focussing the beam from an excimer laser onto a highly oriented pyrolytic graphite (HOPG) target in vacuum or in a low pressure of $H_2$ gas. The focal area on the target is typically $\sim 10^{-2}$ cm$^2$. Each pulse lasts for 30 nsec, and the pulse repetition rate is $\sim$ 15 Hz. The ablation depth on the HOPG surface is $\sim$ 0.1 μm per pulse, corresponding to the removal of $\sim 10^{16}$ carbon atoms per pulse (Duley 1996) This material, in the form of a weakly ionized plasma, impacts on the surface of a quartz, KBr or silicon plate kept at a fixed temperature between 77 and 600 K. The flux of carbon atoms on this surface is $\sim 2 \times 10^{14}$ cm$^{-2}$ per laser pulse. As the surface atom density on the resulting carbonaceous deposit is $\sim 2 \times 10^{15}$ cm$^{-2}$, only 10% of the surface atoms are exposed to new reactive species during each pulse. This means that most reactions will occur at or near the site of each carbon atom impact and only single atom addition reactions will be possible. These reactions will be completed by the time ($\sim$ 50 msec) the next pulse arrives. This combination of deposition conditions, namely single atom reactions followed by a long time interval between the accretion of atoms, corresponds to the situation that would exist when accretion and reaction occurs under interstellar conditions (Duley 2000). This also explains why small rings are preferred at 77 K, since the low temperature of the substrate inhibits the migration of accreted atoms and the sintering of small ring structures to form larger sp$^2$ bonded components. The surface structure shown in Fig. 1b for samples deposited at 77 K is consistent with this scenario, as it shows localized areas of growth separated by voids. Scans at 300 and 600 K show an atomically smooth surface indicating that carbon atoms remain mobile following accretion. This results in the formation of larger $sp^2$ bonded structures, a smaller band gap energy and a shift of $\pi \rightarrow \pi*$ absorption to longer wavelength.

These findings suggest a model in which the carrier of the 217.5 nm band is formed in situ in the DISM via accretion of carbon atoms and ions from the interstellar gas. Given the abundance of silicate grains, it is logical to assume that this accretion happens on the surface of these particles. This supports the traditional view of "core-mantle" grains in the DISM (Whittet 2003), but would seem to be inconsistent with the properties of the 217.5 nm band that require it to be produced by absorption in nano-particles, rather than as an extinction feature arising in larger grains (Draine 1998). It also does not agree with observations showing that the 3.4 $\mu$m IR absorption band is un-polarized along lines of sight through the DISM where the 10 $\mu$m silicate feature is polarized (Mason et al. 2007). However, both of these conclusions are based on grain models in which particles are represented by smooth geometrical spheroidal figures with an even distribution of oscillators inside the material. A more realistic physical picture for the structure of core-mantle grains would be one in which the silicate core is porous or layered and accretion of carbon from the gas phase results in the selective growth of nano-carbon particles supported by this framework. A simple representation of one type of such a structure is shown in Fig. 11. It is evident that the mass distribution of carbon nano-particles within this composite need not replicate that of the silicate framework, implying that dichroic polarization by the silicate would not necessarily be reproduced by the carbon component. Carbon nano-particles supported in this way should also produce a 217.5 nm absorption band as this configuration is basically the same as that appearing in Fig. 1b where nano-particles are collected on a silica surface and the spectrum (Figs. 7 & 8) shows this feature. Recent scattering calculations (Iati et al. 2011) using a stratified silicate-carbonaceous grain structure with porosity has reproduced the wavelength dependence of the interstellar extinction curve throughout the UV.

The IR bands at 3.2, 6.2, 7.7, 8.6, & 11.3 μm, associated with aliphatic and aromatic hydrocarbons, are also seen in emission from the DISM (Sakon et al. 2004). The PAH model attributes this emission to individual molecules excited by photons from the interstellar radiation field, but this interpretation has recently been questioned as it appears that these bands can also be excited thermally by chemical energy liberated on the recombination of stored hydrogen and other reactive species (Duley & Williams 2011). This recombination, producing transient heating to temperatures in excess of 1000 K, has been well documented in laboratory experiments (Wakabayashi et al. 2004, Yamaguchi & Wakabayashi 2004) and is quite likely to occur in structures such as that shown in Fig. 11. In this case, the energy source would involve the recombination of stored H atoms (Sugai et al. 1989), but radical recombination is also possible given the heterogeneous composition of these carbonaceous nano-particles (Biener et al. 1994, Jariwala et al. 2009). Laboratory experiments have demonstrated that thermal energy released from an amorphous carbon layer on a core-mantle grain can produce significant heating (Kaito et al. 2007). The overall effect would be to excite the 3.2, 6.2, 7.7, 8.6, & 11.3 μm bands in emission from the carbon nano-particles as shown in Fig. 11. Localization of this chemistry to the nanoparticles on the surface of the silicate would ensure that emission occurs only from the carbonaceous component.

## 7. FORMATION MECHANISM

Spectroscopic analysis of deposited samples indicates that these materials contain both chains and rings and that hybrid structures consisting of chains connected to rings are also present. The fraction of carbon in the ring component is relatively small in the 77 K deposits and increases at deposition temperatures above 300 K. As materials are deposited on an atom by atom basis, and H atoms are always present in diffuse clouds, the initial structures are likely created by C atom addition to ethynyl radicals, $C_2H$, forming propynylidyne, $C_3H$, butadiynyl, $C_4H$, and longer acetylenic chains, $C_nH$. Various isomers of these radicals exist, including the cyclic structures c-$C_3H$, and c-$C_5H$. Longer chains such as the hexatriynyl radical, $C_6H$, can contain C=C as well as C≡C groups, and these chains are unstable with respect to the formation of phenyl, Ph-$(C≡C)_nH$, and diphenyl, Ph-$(C≡C)_n$-Ph, compounds, where Ph is the phenyl group, $C_6H_5$. Raman spectra (Fig. 5) indicate that long polyyne and cumulene chains and phenyl derivatives including diphenylacetylene and diphenylbutadiene (table 2) are present in HAC nanoparticles, while hydrogenation of these compounds will result in the formation of olefinic and aliphatic hydrocarbon chains giving rise to the $CH_n$ features seen in the spectrum. Intermediate components would include polyacetylenic chains, $C_nH_{n+2}$ where n = 2, 4,.. Branched hydrocarbon chains are evident in the spectra shown in Fig. 6.

A collection of reactive species such as the acetylenes and related components on the surface of a solid will be stable only up to a critical concentration. This concentration will be determined by the geometry and the temperature of the substrate as well as by the dimensions and composition of the reactants. When the critical concentration is exceeded, then reactive species are physically close enough that they can combine. A well known reaction that occurs when acetylenic groups are present on a surface is the trimerisation $3C_2H_2 \rightarrow C_6H_6$ in which benzene is formed from three acetylene molecules. This reaction is highly exothermic and the energy liberated as heat from this type of reaction may contribute to the excitation of emission from HAC nanoparticles as discussed above.

A number of spectral features in these samples can be tentatively associated with naphthalene and naphthalene derivatives. A straightforward chemical route to the formation of naphthalene under the conditions of the current experiments would involve the sequential addition of an ethynyl group, $C_2H$, to phenylacetylene followed by rearrangement to form a naphthyl radical, $C_2H + (C_6H_5)C_2H \rightarrow C_{10}H_7$. This enables the formation of a variety of naphthalene derivatives including binaphthyl, $(C_{10}H_7)_2$ whose presence is indicated in Raman spectra (table 2).

The model that is indicated from this analysis is that of the in situ formation of hydrogenated amorphous carbon nanoparticles by accretion of individual carbon atoms from the gas phase under DISM conditions. This accretion occurs on the surface of irregular porous silicate particles and initially leads to the formation of small chains. Above a certain length, these chains become unstable with respect to cyclic structures, particularly to the formation of phenyl. Reactions between radicals also promote ring formation, and as both of these effects occur, the composition evolves into a mixture of ring and chain compounds. Phenyl and naphthyl radicals act as key templates for the formation of a variety of substituted rings and larger structures, but the chemistry continues to be driven by addition reactions involving ethynyl groups being created on the surface by the continued accretion of incident carbon atoms from the ambient gas. This will result in HAC particles whose internal composition is enriched in small aromatic rings, and whose outer surface contains a higher concentration of chains and radicals. The chemical energy associated with reactions between radicals and with the recombination of hydrogen atoms trapped within this material is periodically released leading to heating of the particle and to emission of the IR spectral features associated with aromatic hydrocarbons.

## 8. DUST IN DENSE CLOUDS

While the aliphatic $CH_n$ absorption bands near 3.38, 3.42, and 3.48 μm (2960, 2925 & 2875 cm$^{-1}$, respectively) are detected in the DISM in Galactic and extragalactic sources, these spectral features disappear in the absorption spectrum of dense cloud material and are replaced by a broad absorption band at 3.469 μm (2882 cm$^{-1}$) with a FWHM of 0.105 μm (87 cm$^{-1}$) (Brooke et al. 1999). As noted in section 3, this feature is characteristic of the stretching mode of tertiary CH groups found in many hydrocarbons, but most notably on the surface of diamond. Analysis of the 3.4 μm band in DISM dust (Dartois et al. 2007) reveals a feature at 3.476 μm (2877 cm$^{-1}$), but assignment of the interstellar at 3.47 μm to tertiary CH on sp$^3$ bonded carbons is complicated by an overlap with Fermi resonance features at the same wavelength (Table 7). This implies that a 3.47 μm feature detected along with the absorption bands of $CH_2$ and $CH_3$ can most likely be assigned to a Fermi resonance, rather than to tertiary CH. Conversely, the appearance of a 3.47 μm feature in the absence of $CH_2$ and $CH_3$ is indicative of a dust component containing tertiary CH.

The observation of a 3.47 μm band in dense clouds is then consistent with a change in dust composition whereby the concentration of $CH_2$ and $CH_3$ groups is reduced, but carbon dust is otherwise the same as that existing in the DISM. In practice, this could occur if aliphatic chains are converted to acetylenic or cumulenic chains or are completely eliminated, while the remaining material retains the original concentration of diamond-like carbon. Dehydrogenation of aliphatic chains in dense cloud dust could be produced by heating following radical reactions, through absorption of residual UV radiation or chemically via H atom reactions (Mennella 2010). Overall, our analysis indicates that diamond-like carbon inclusions may be present in both DISM and dense cloud material as evidenced by the appearance of absorption attributable to tertiary CH in both types of environment.

## 9. CONCLUSIONS

Samples of hydrogenated amorphous carbon prepared by single atom deposition of carbon at 77 K results in the formation of nano-particles whose spectroscopic properties include the appearance of a 217.5 nm absorption band. The characteristics of this band reproduce those of the interstellar feature as catalogued by Fitzpatrick & Massa (1986). Chemical analysis of this material obtained from XPS, Raman and SERS spectra show that it consists of a network of chains and rings. The ring component is limited to small $sp^2$ bonded molecules, including a variety of naphthalene derivatives, which we suggest are the source of the 217.5 nm absorption band. The spectra contain evidence for acetylenic, aliphatic and olefinic chains containing up to six carbon atoms, as well as hybrid structures involving chains bonded to rings. One of these is diphenylacetylene. Compositions that result in the enhancement of the 217.5 nm band correspond to an $sp^2$ bonded content of ~ 0.3 and an $sp^3$ bonded content of ~ 0.7, implying that deposition under these conditions tends to form carbon chains, rather than rings. We discuss this mechanism in some detail and outline its implications for the formation, excitation and evolution of HAC nano-particles in the interstellar medium.   We conclude that HAC is formed in situ on the surface of silicate grains in the DISM, and suggest a simple grain model that incorporates this structure.

   This material has a rich vibrational spectrum that can be resolved into a large number of components corresponding to specific modes of the chemical structures contained in these nanoparticles. This spectroscopic data has been used to produce detailed fits to emission in type A and B sources as determined by van Diedenhoven et al. (2004) as well as that of the PPN IRAS 01005+7910 (Zhang & Kwok 2011).  We believe that these represent the first accurate fits to astronomical spectra of this kind using purely experimental measurements on carbonaceous materials produced in the laboratory.  Further investigation of the properties of these materials is warranted, as they would seem to represent an accurate simulation of interstellar carbon dust.

This research was supported by grants from the NSERC of Canada

Table 1. Energies of spectral components from Raman spectra of samples deposited at 77, 300 and 600 K (Fig. 2).  Raman spectra were excited with 630 nm radiation.

Deposition temperature (K)

| 77 | | 300 | | 600 | |
|---|---|---|---|---|---|
| Energy | FWHM | Energy | FWHM | Energy | FWHM |
| 1551 cm$^{-1}$ | 155 cm$^{-1}$ | 1597 cm$^{-1}$ | 76 cm$^{-1}$ | 1595 cm$^{-1}$ | 93 cm$^{-1}$ |
| 1450 | 125 | 1531 | 96 | 1478 | 145 |
| 1331 | 330 | 1362 | 132 | 1339 | 144 |
| | | 1256 | 175 | 1219 | 144 |
| | | | | 1106 | 270 |

Table 2a. SERS features in carbonaceous deposits in the region of the G band. All energies are in cm$^{-1}$. The most intense feature is highlighted*. Due to SERS enhancement both IR and Raman bands appear in these spectra and the relative intensity of individual features can vary. Some representative simple molecules with spectral features at these energies are: N (naphthalene), BN (binaphthyl), (EN) ethylnaphthalene, (HD) hexadiene, (PD) pentadiene, (PN) phenanthrene, (MPN) methylphenanthrene, (DMPN) dimethylphenanthrene

| Deposition temperature (K) | | | |
|---|---|---|---|
| 77 | 300 | 600 | |
| Energy | Energy | Energy | Assignment |
| | | 1698 | C=C olefinic |
| | | 1669 | C=C olefinic, eg. PD |
| | | 1647 | C=C olefinic, eg. HD |
| 1617 | | 1626 | sp$^2$ ring, EN, MPN, DMPN |
| 1607 | 1603* | 1601* | sp$^2$ ring, EN, DMN, PN, MPN |
| 1591* | | 1585 | sp$^2$ ring, BN |
| 1578 | | | sp$^2$ ring, N |
| 1562 | 1563 | 1561 | sp$^2$ ring, N, BN |
| | | 1529 | DMPN |
| 1512 | | 1512 | N, DMN |
| 1502 | 1500 | 1502 | N, EN, DMN, PN, MPN |

Table 2b. SERS features in carbonaceous deposits between 1440 and 1000 cm$^{-1}$. All energies are in cm$^{-1}$. Due to SERS enhancement both IR and Raman bands appear in these spectra and the relative intensity of individual features can vary. The most intense feature is highlighted*. Molecules are: N (naphthalene), DPA (diphenylethylene), PB (phenylbutene), MN (methylnaphthalene), EN (ethylnaphthalene), DMN (dimethylnaphthalene), BN (binaphthyl), DPE (diphenylethylene), DPB (diphenylbutadiene), OT (o-terphenyl), BP (biphenyl), DPH (diphenylhexadiene), PPB (propenylbenzene), DPA (diphenylacetylene), PN (phenanthrene), MPN (methylphenanthrene), DMPN (dimethylphenanthrene), TP (triphenylene).

Deposition temperature (K)

| 77 | 300 | 600 | |
|---|---|---|---|
| Energy | Energy | Energy | Assignment |
| 1467 | | 1475 | $CH_n$, PN, MPN |
| | 1441 | 1443 | $CH_n$, N, PN, MPN |
| 1432 | | 1424 | $CH_n$, PN, MPN, TP |
| 1413 | 1409 | 1401 | DPE, PB |
| 1377 | 1378 | 1377 | $CH_n$, N, MN, BN, MPN, DMPN |
| 1360 | | | $CH_n$ |
| 1343 | 1348 | 1347 | $CH_n$, DPE, DMPN |
| 1329 | 1327 | 1322 | Nanodiamond, DPE |
| 1305 | 1311 | 1310 | $CH_n$. PN, MPN |
| | 1298* | 1294* | DPB, OT, DMPN |
| 1287 | 1286 | | |
| 1275* | | 1274 | N, MPN, BP, DPH |
| 1264 | 1267 | | EN, BN, DPH |
| 1252 | 1241 | 1244 | N, PN, MPN, DMPN, TP |
| 1238 | 1228 | 1229 | PN |
| 1216 | 1214 | 1208 | DMN, PPB |
| 1193 | | | DPB, DPE |
| | | 1178 | DPB |
| 1162 | 1166 | 1156 | BP, PPB, DPB, TP |
| 1140 | 1136 | | DPB, DPA, PN, MPN, TP, $CH_n$ |
| 1117 | | 1121 | |
| 1100 | 1093 | 1084 | PN |
| 1069 | 1040 | | $CH_n$, subst-benzene, PN |
| 1002 | 1004 | | subst-benzene |

Table 3.  SERS features between 1000 and 350 cm$^{-1}$ in carbonaceous deposits.  Energies are in cm$^{-1}$.  Molecular designations are as given in Table 2.

Deposition temperature (K)

| 77 | 600 | |
|----|-----|---|
| Energy | Energy | Assignment |
| 967 | | DMN, CH$_n$ |
| 927 | 916 | CH$_n$ |
| | 879 | solo CH |
| 865 | | CH$_n$, MN, BN |
| 855 | 848 | duo CH, CH$_n$, DMPN |
| 822 | | DMN, PN |
| 790 | 788 | N, OT, trio CH |
| 768 | | N, DMN, EN, trio CH, CH$_n$ |
| | 743 | DMN, BP, PN, TP, quartet CH |
| 722 | 720 | DPE, OT |
| 685 | 680 | BN |
| | 667 | MN |
| 620 | | EN, DPP, OT, PN, TP |
| 598 | 589 | DPE |
| 527 | | DMN, BN, DMPN |
| | 538 | DMN, BN, DMPN |
| | 510 | N, MN, EN, BN |
| 485 | 482 | N, MN |

| 428 | 439 | DMPN |
| | 408 | DMN, OT |
| 386 | 395 | DPA |
| | 366 | OT |

Table 4. Wavelength, $\lambda_m$, of the most intense absorption band in individual molecular components for $\lambda > 200$ nm. The designation of molecules is as given in Table 2. Data is from the NIST WebBook.

| Molecule | $\lambda_m$(nm) |
|----------|-----------------|
| DPE | 215 |
| N | 218 |
| PD | 220 |
| OT | 225 |
| PA | 235 |
| BP | 245 |
| PN | 250 |
| MPN | 250 |
| TP | 258 |
| DPA | 278 |
| DPB | 285, 334 |
| BD | 290 |
| DPB | 332 |

Table 5. Spectral components appearing in the fit to the spectrum of the PPN IRAS 01005+7910 (Zhang & Kwok 2011) (Fig. 9). These components all appear in measured spectra of laboratory samples (Table 2).

| Wavelength (μm) | Energy (cm$^{-1}$) | FWHM (cm$^{-1}$) |
|---|---|---|
| 6.22 | 1608 | 20 |
| 6.40 | 1563 | 20 |
| 6.61 | 1512 | 30 |
| 6.82 | 1467 | 30 |
| 7.02 | 1424 | 15 |
| 7.26 | 1377 | 30 |
| 7.35 | 1360 | 20 |
| 7.45 | 1342 | 20 |
| 7.56 | 1322 | 30 |
| 7.91 | 1264 | 35 |
| 8.08 | 1238 | 20 |
| 8.24 | 1214 | 25 |
| 8.38 | 1193 | 20 |
| 8.58 | 1166 | 20 |

Table 6. Spectral components appearing in the fits to Type A and Type B emission spectra (van Diedenhoven et al. 2004). These components are seen in spectra of laboratory samples (Table 2).

| | Type A | | | Type B | |
| Wavelength | Energy | FWHM | Wavelength | Energy | FWHM |
| μm | cm$^{-1}$ | cm$^{-1}$ | μm | cm$^{-1}$ | cm$^{-1}$ |
|---|---|---|---|---|---|
| 7.446 | 1343 | 15 | 7.446 | 1343 | 15 |
| 7.524 | 1329 | 20 | | | |
| 7.564 | 1322 | 15 | 7.564 | 1322 | 15 |
| 7.628 | 1311 | 10 | 7.628 | 1311 | 10 |
| | | | 7.662 | 1305 | 20 |
| 7.728 | 1294 | 20 | 7.728 | 1294 | 20 |
| 7.843 | 1275 | 15 | | | |
| | | | 7.874 | 1270 | 20 |
| | | | 8.039 | 1244 | 15 |
| 8.382 | 1193 | 10 | | | |
| 8.606 | 1162 | 20 | 8.621 | 1160 | 15 |
| | | | 8.681 | 1152 | 15 |

Table 7. Characteristics of the 3.47 μm band observed in dense clouds (Brooke et al. 1999) and the DISM (Dartois et al. 2007) compared with laboratory data for samples prepared at 77 and 120 K.  HAC data is from Grishko & Duley (2000).

| Wavelength, μm (energy, cm$^{-1}$) | FWHM, μm (cm$^{-1}$) | Object |
| --- | --- | --- |
| 3.469 (2882) | 0.105 (87) | Dense cloud |
| 3.476 (2877) | 0.048 (40) | DISM |
| 3.470 (2882) | 0.048 (40) | Fig. 6, 77 K |
| 3.478 (2875) | 0.073 (60) | 120 K |
| 3.465 (2886) | 0.145 (121) | HAC, 300 K |

FIGURE CAPTIONS

Fig. 1. SEM scans of samples deposited at 300 K (a) and at 77 K (b). The deposit at 300 K, which has the same morphology as that of deposits at 600 K, is smooth and exhibits none of the nano-structure that appears in that at 77K. We attribute this to a combination of higher surface mobility for atoms deposited at 300 K together with thermal annealing of small $sp^2$-bonded clusters at elevated temperature. The structure at 77 K arises from the clustering of nano-scale clusters formed sequentially by deposition of individual atoms. The deposition conditions are such that each site on the growing surface sees less than one atom per laser pulse. Limited surface mobility of accreted carbon atoms at 77 K favors the growth of small (ring and chain) $sp^2$ bonded structures (see section 6).

Fig. 2. Raman spectra of samples deposited at 77, 300 and 600 K. All spectra were recorded at 300 K at an excitation wavelength of 632.8 nm. These spectra show broad "D" and "G" bands near 1330 and 1600 $cm^{-1}$, respectively. The D band arises from aromatic rings while the G band is due to $sp^2$ bonded carbon in chains and rings.

Fig. 3. Spectra recorded under SERS conditions for samples prepared at 77, 300 and 600 K. All spectra were measured at 300 K. Both IR and Raman active transitions appear in this spectrum. Suggested assignments of individual spectral features are given in Table 2.

Fig. 4. SERS spectra of samples at energies < 1000 $cm^{-1}$, for samples deposited at 77, 300 and 600 K. Spectral features in this region correspond to bending and deformation modes of individual chemical groups and include features that can be associated with the bending vibration of CH groups on aromatic rings.

Fig 5. SERS spectra of samples showing features arising from polyyne and cumulene chains. The feature at 2233 cm$^{-1}$ in the 77 K sample may indicate the presence of diphenylacetylene. Other bands at 2087, 2113, 2130 and 2233 cm$^{-1}$ in the 77 K samples are due to polyyne chains containing up to 10 carbon atoms.

Fig 6. SERS spectra of samples showing CH, CH$_2$ and CH$_3$ stretching bands (3100-2800 cm$^{-1}$) and a variety of CH$_n$ overtone bands (2800-2500 cm$^{-1}$). The feature at 2882 cm$^{-1}$ is a Fermi resonance, but could also indicate the present of tertiary CH groups in nano-diamond inclusions. The band at 2722 cm$^{-1}$ can be identified with dimethylnaphthalene. Other bands at 2780, 2700 and 2615 cm$^{-1}$ are characteristic of specific aliphatic bond configurations (see text for details).

Fig. 7. UV absorption spectra of samples prepared at 77, 120, 300 and 600 K. Each spectrum is the result of overlapping absorption bands corresponding to $\pi \rightarrow \pi^*$ transitions in sp$^2$ bonded carbons in specific chemical groups (table 4). The absorption peak near 220 nm in the spectrum of the 77 K deposit is tentatively assigned to naphthalene and some simple naphthalene derivatives. The vertical line with long dashes corresponds to the central wavelength of the interstellar absorption band, while the short-dashed vertical lines indicate the FWHM of the band.

Fig. 8. Spectral profile of 217.5 nm components found in absorption in different samples. Variations in the central wavelength, energy ($\lambda_0^{-1}$), and FWHM ($\gamma$) of these bands can be attributed to slight changes in local bonding. a) Sample prepared at 77 K and measured at 300 K. $\lambda_0^{-1} = 4.57$ μm$^{-1}$ and $\gamma = 0.8$ μm$^{-1}$ (5.67 eV and 0.99 eV respectively). There is another weak feature, likely due to another type of molecular group, present between 3.8-4.0 μm$^{-1}$. b) Sample prepared at 77 K in 0.1 Torr of H$_2$ gas and measured at 300 K. $\lambda_0^{-1} = 4.57$ μm$^{-1}$ and $\gamma = 1.28$ μm$^{-}$

[1] (5.67 eV and 1.59 eV respectively). There is another weak feature between 4.4 and 4.2 μm-1 (5.45 and 5.21 eV) in this spectrum.  c) Sample prepared at 300 K in vacuum.  $\lambda_0^{-1}$ = 4.63 μm$^{-1}$ and γ = 1.0 μm$^{-1}$ (5.74 eV and 1.24 eV respectively).

Fig 9.  Emission spectrum of the PPN IRAS 01005+7910 (Zhang & Kwok 2011), together with a fit to this spectrum using laboratory spectra.  The parameters of this fit are summarized in table 5.

Fig 10.  Prototypical Type A and Type B emission in the middle IR as reported by van Diedenhoven et al. (2004) with fits derived from laboratory spectra of samples produced here. Fit parameters are listed in table 6.  There is a clear distinction between these fits, indicating that the material responsible for this emission is different in the two types of source.  The type A emission spectrum in this region is that of IRAS 233133+6050, while the type B spectrum is that of HD 44179.

Fig. 11.  Schematic cross-section of a portion of layered silicate (rectangles) and carbon nano-particles (spheres and spheroids) formed on the surface of the silicate by accretion of carbon atoms and ions from the gas phase.  The 217.5 nm feature and hydrocarbon absorption/emission would be produced by the carbon nano-particles, while the silicate is responsible for a 9.7 μm band which is polarized by dichroic absorption.  The mass distribution in the carbon nano-particle component does not replicate that of the silicate and so the 217.5 nm band, together with IR absorption bands from HAC nano-particles are not polarized.

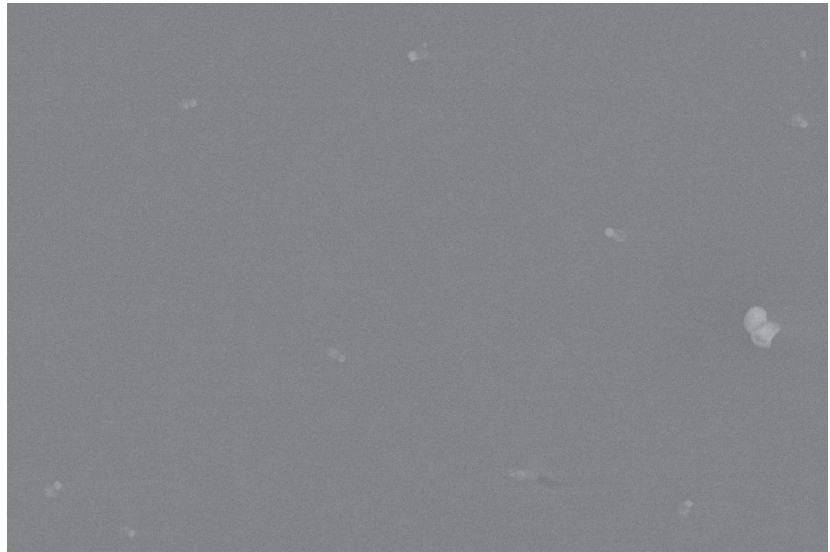

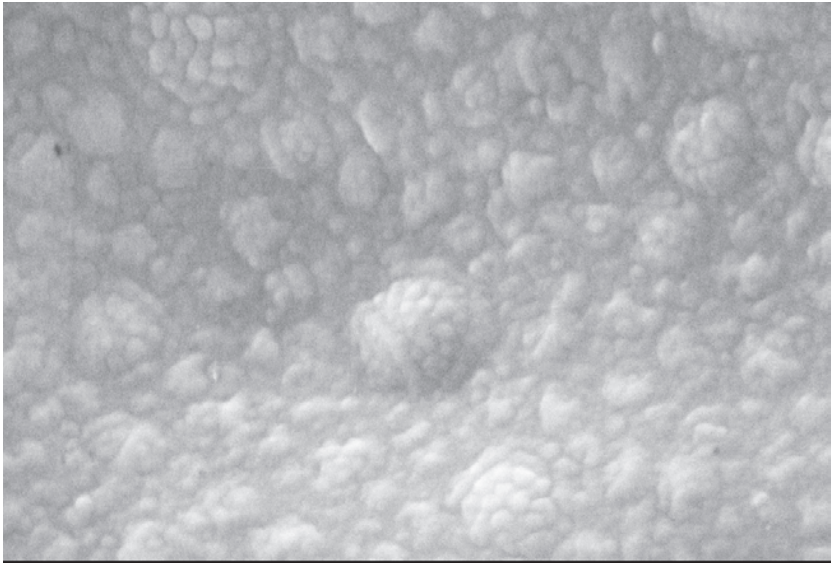

Width = 4.145 µm     100nm     Mag = 50.00 K X    WD = 7 mm    Date :16 May 2005   Time :15:23:09
File Name = sample4-2-50kx.tif         EHT = 5.00 kV    Signal A = InLens   System Vacuum = 1.28e-006 mBar
Waterloo Advanced Technology Laboratories - www.WATLabs.com     User Name = OS DULEY      University of Waterloo   LEO FESEM 1530

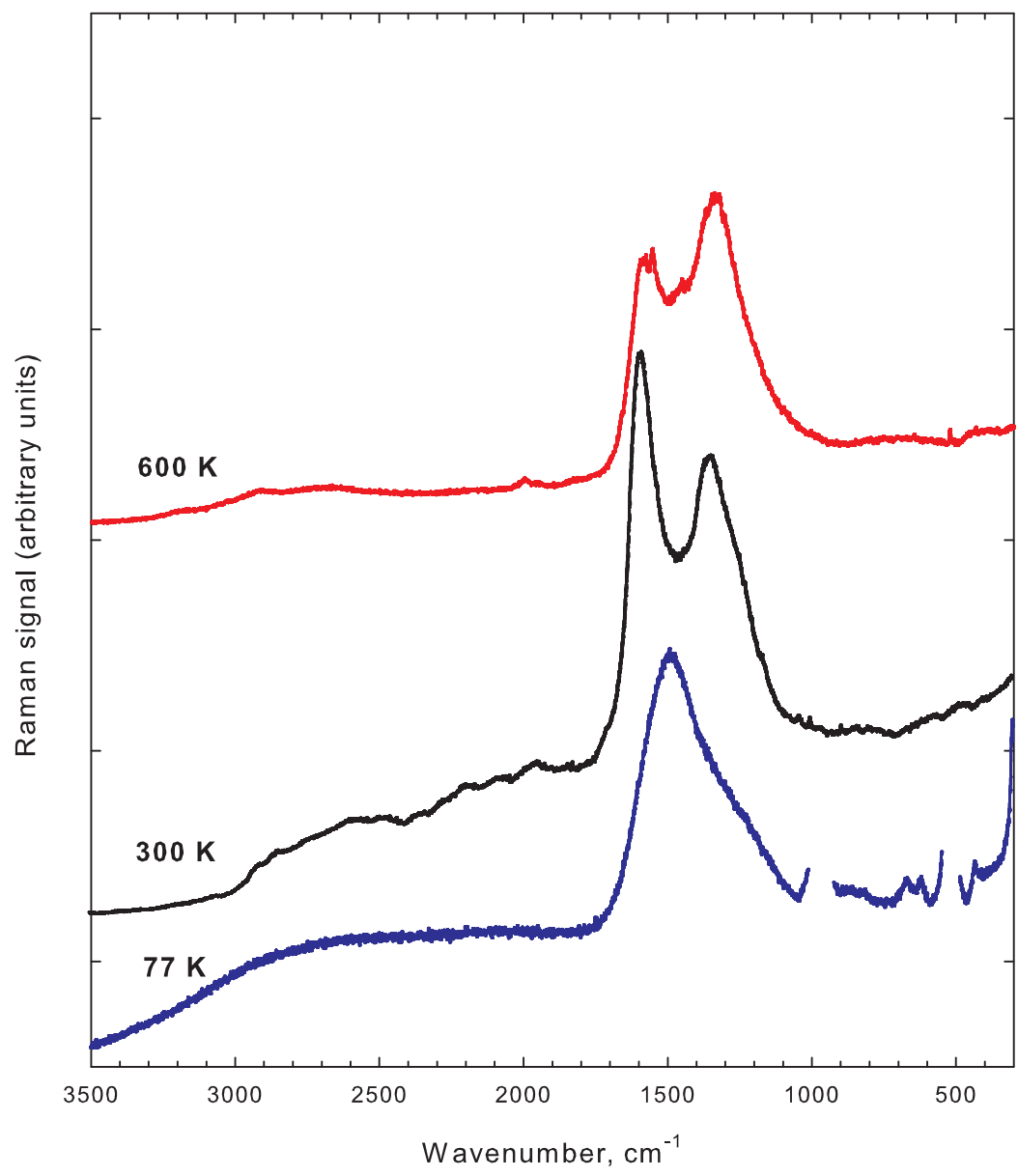

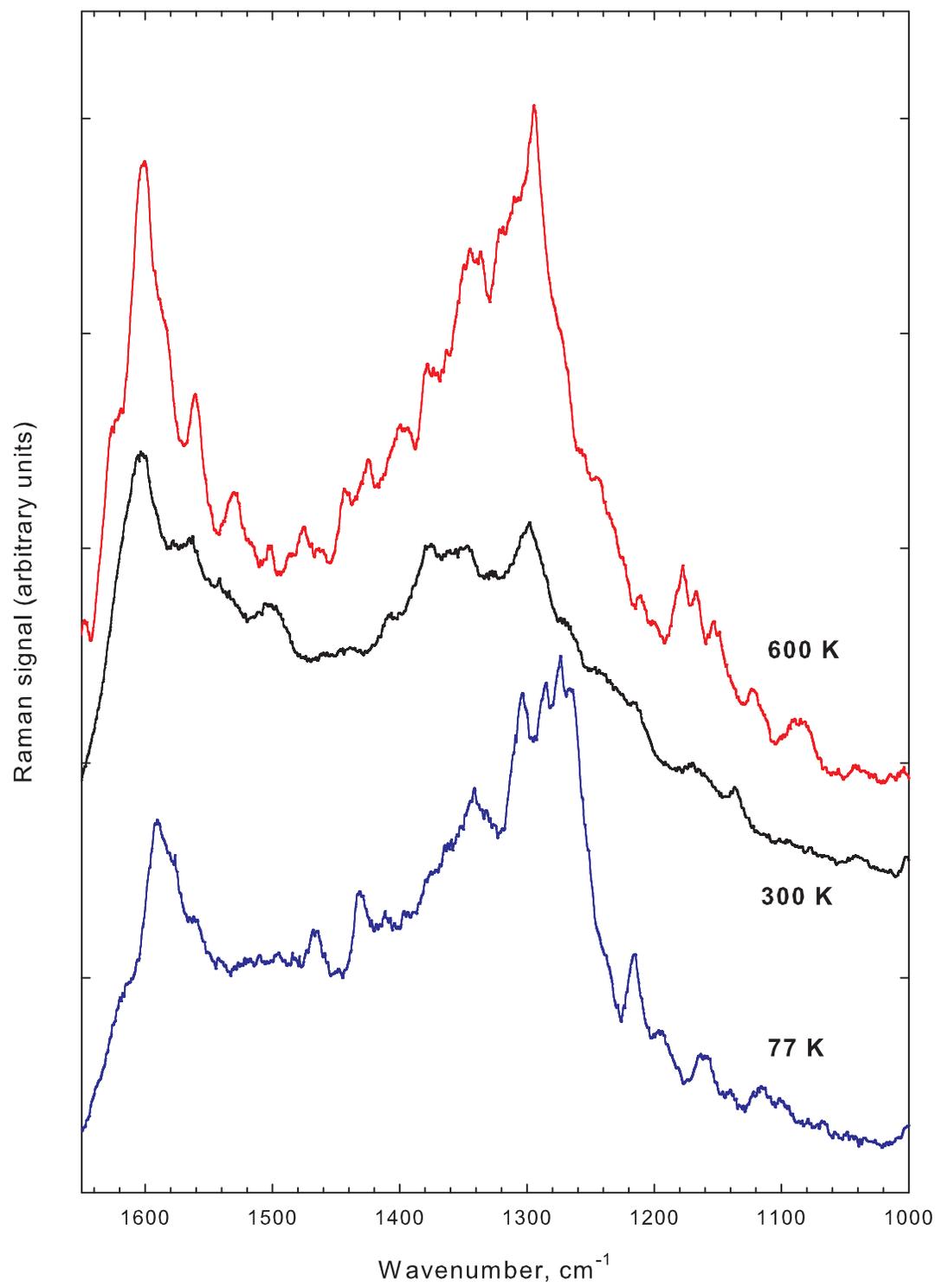

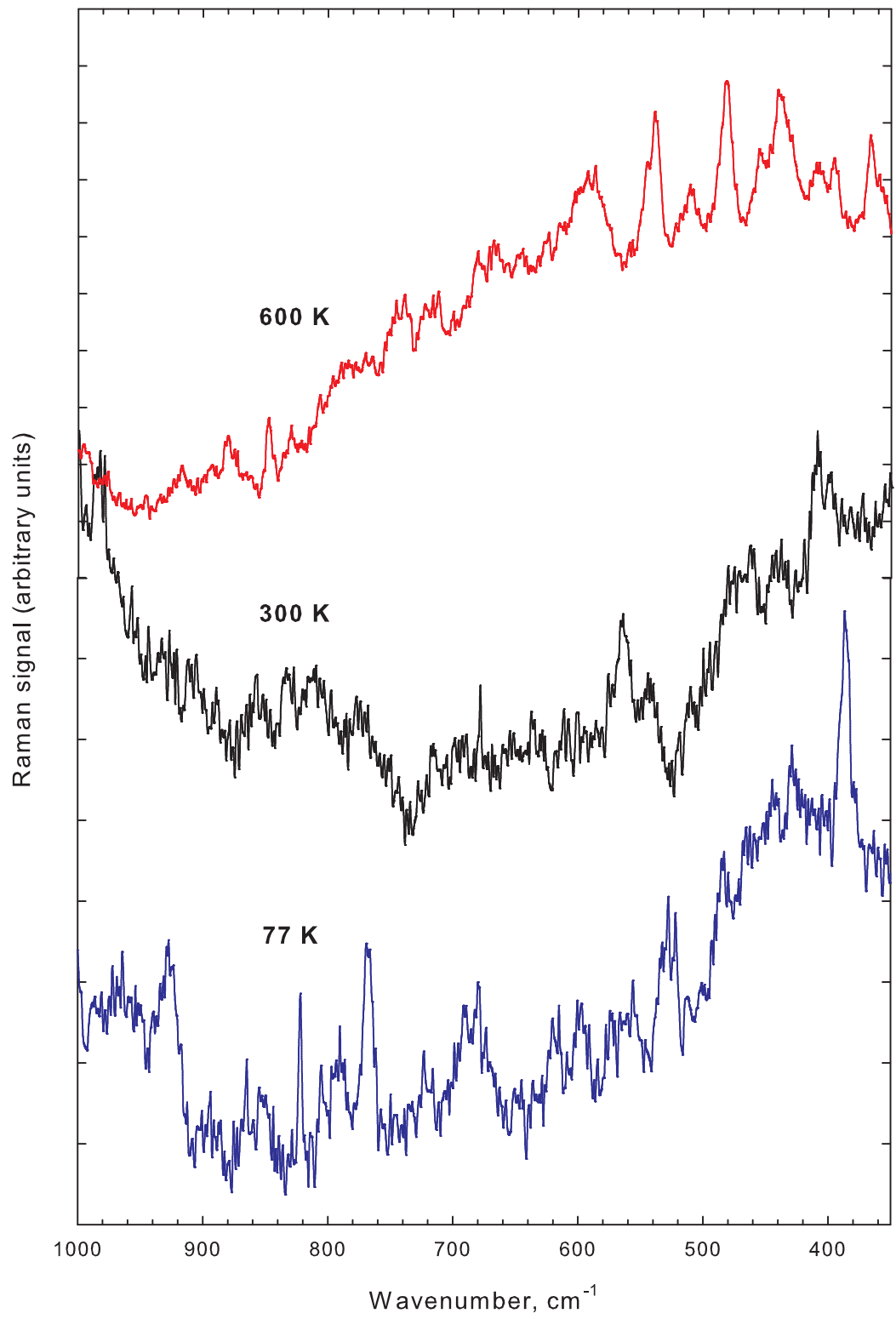

**600 K**

**300 K**

**77 K**

Raman signal (arbitrary units)

Wavenumber, cm⁻¹

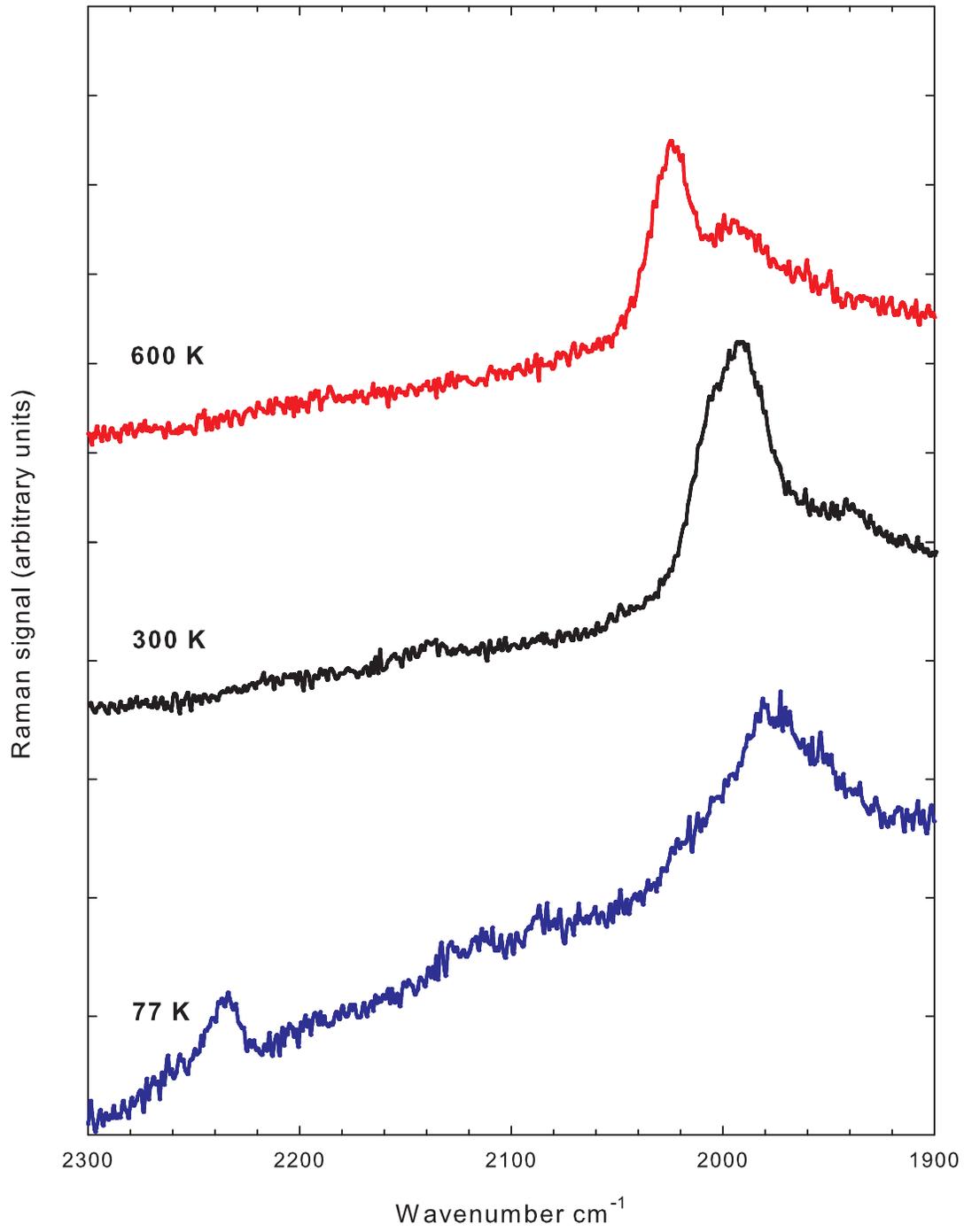

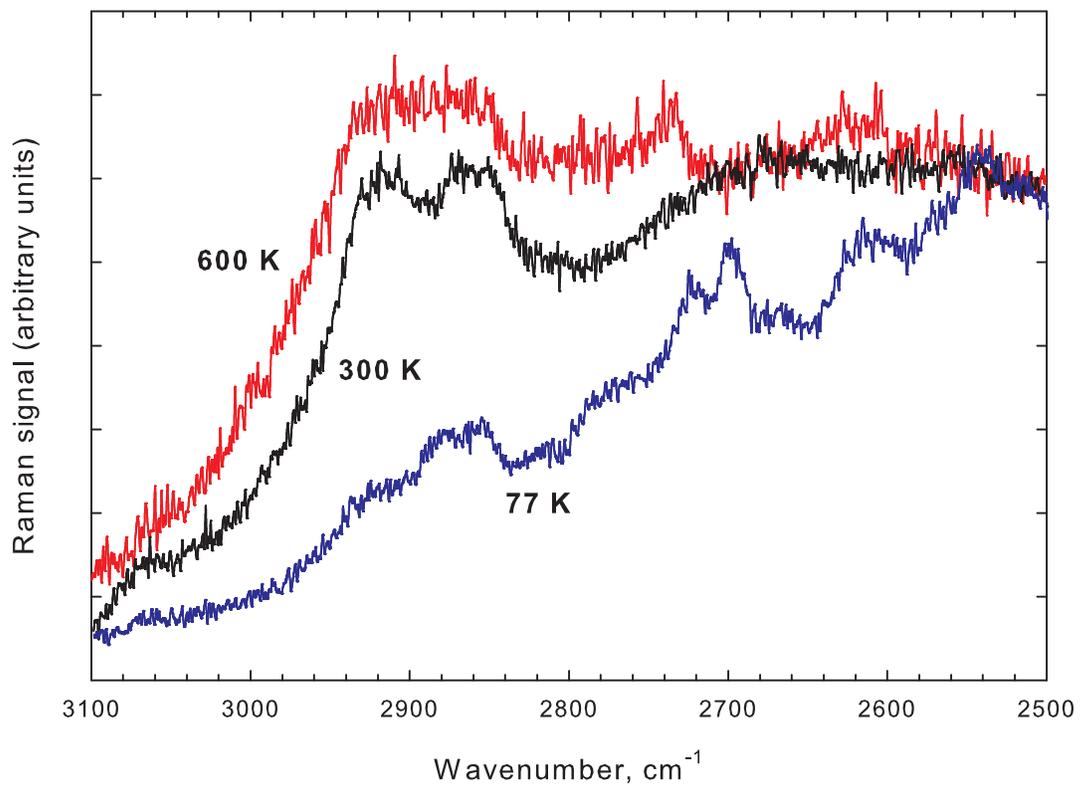

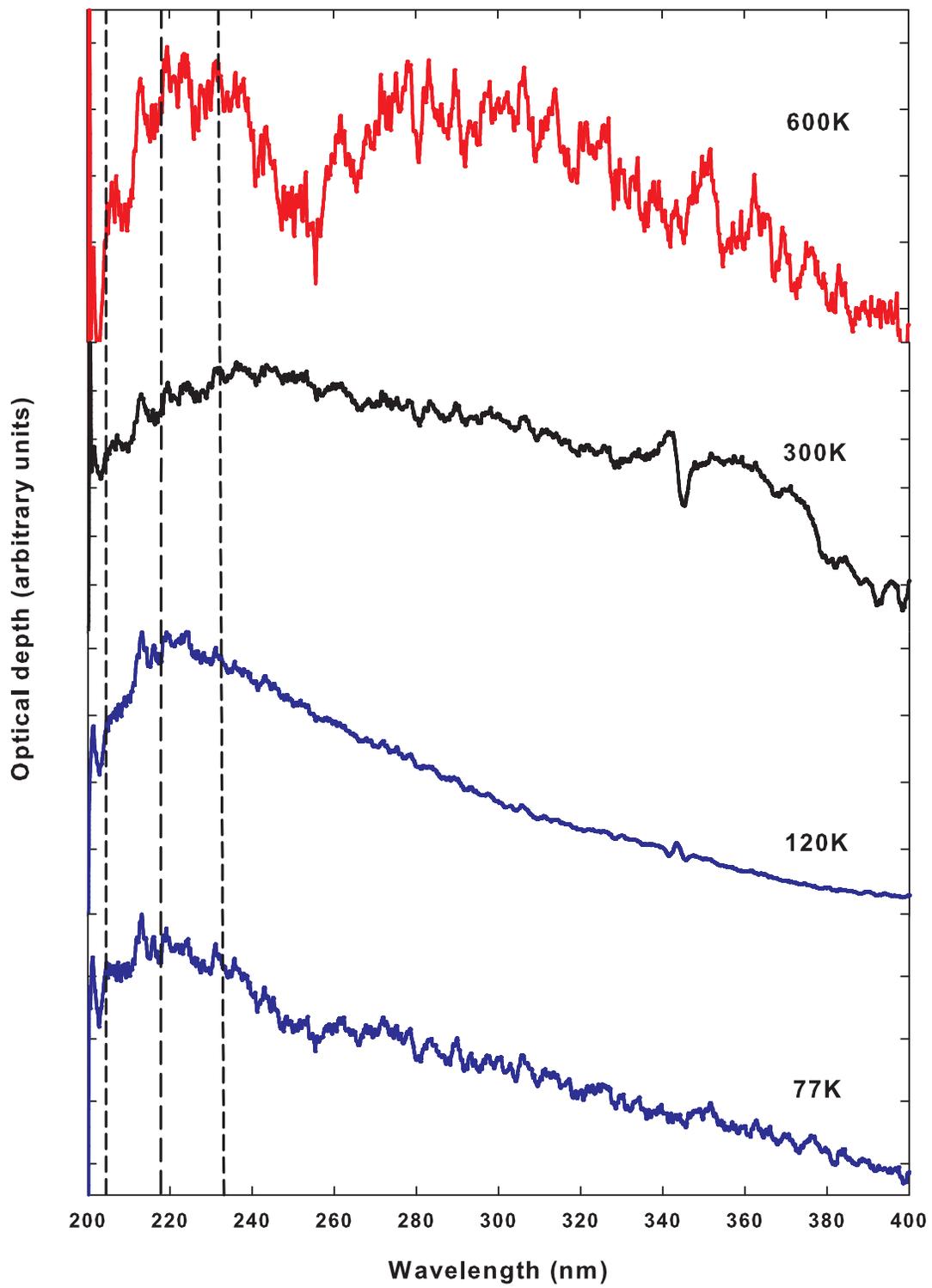

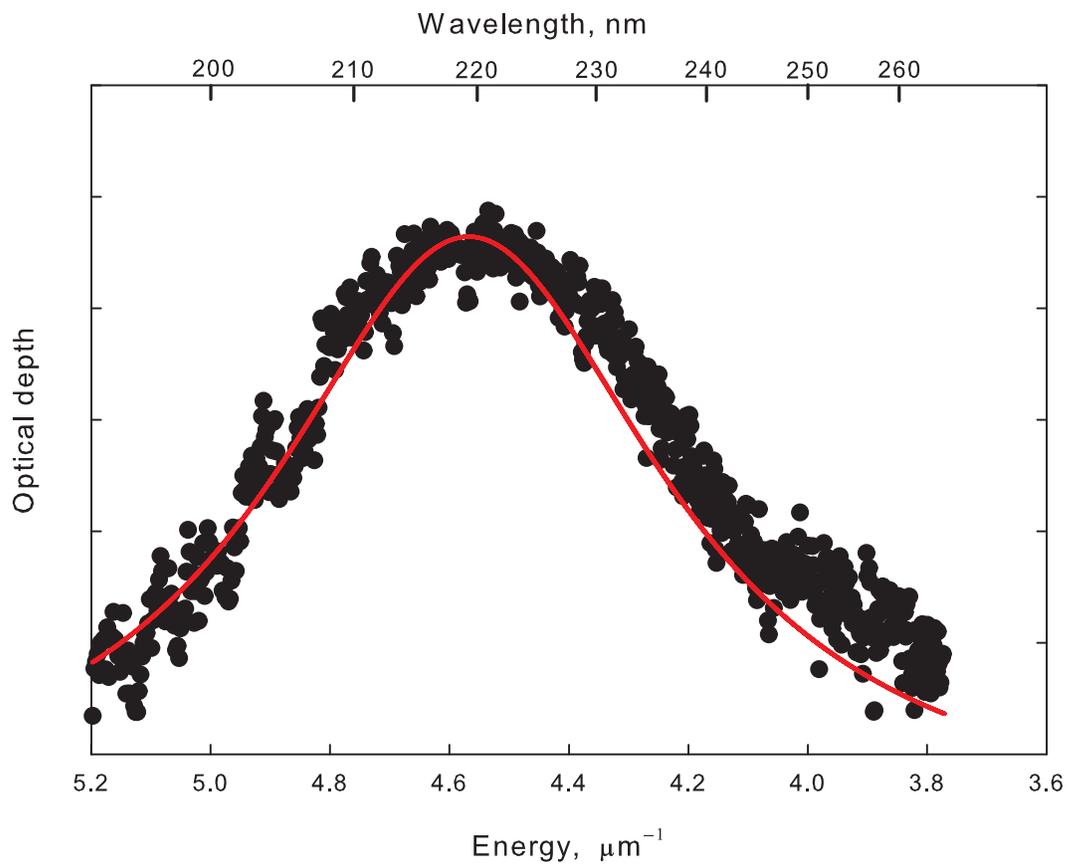

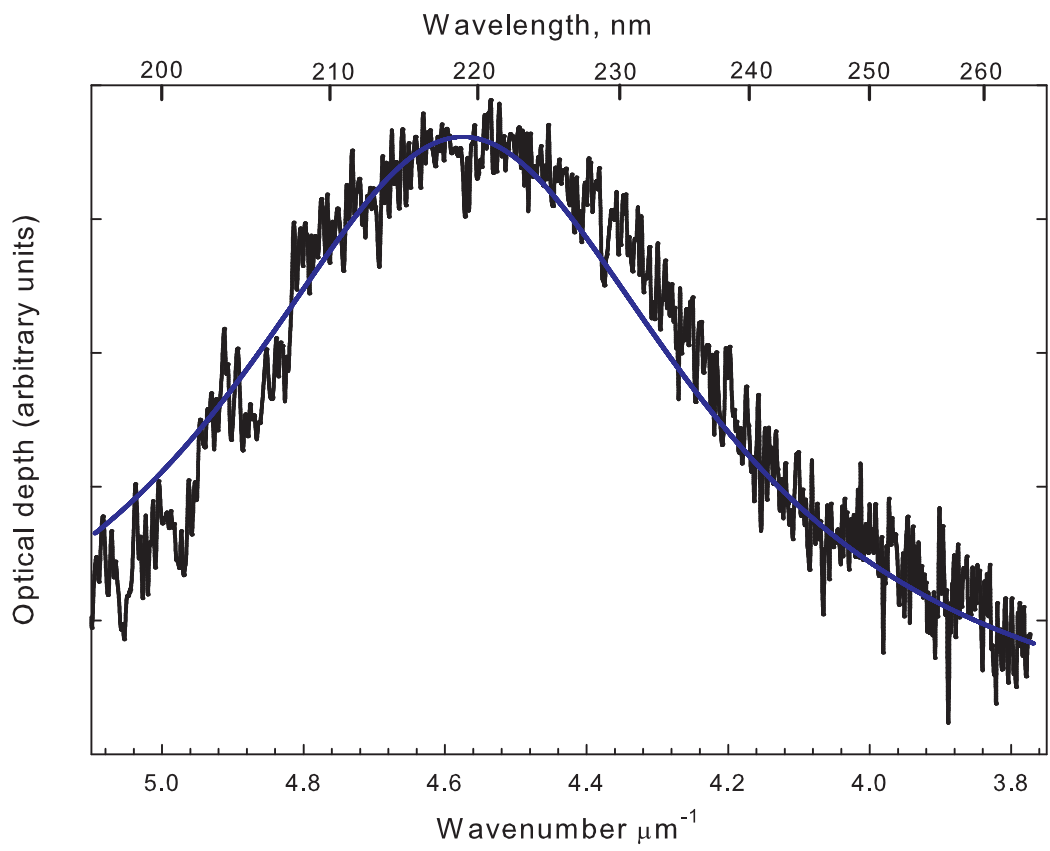

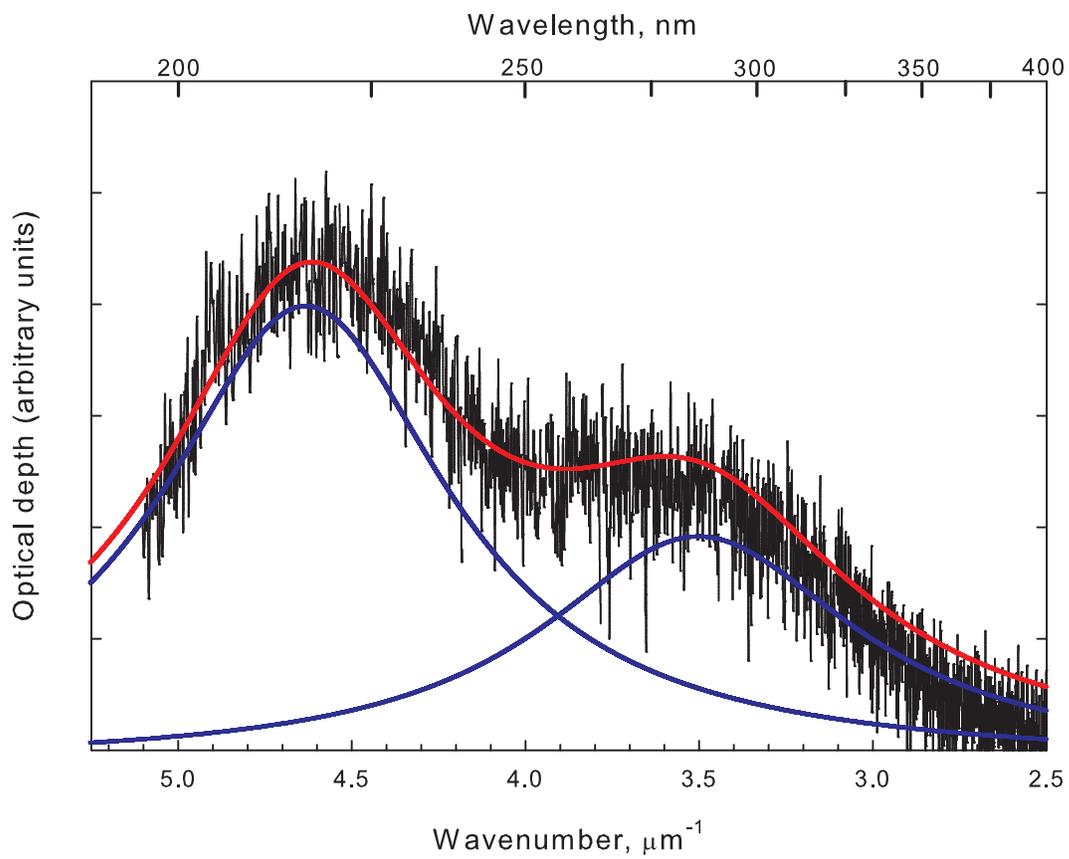

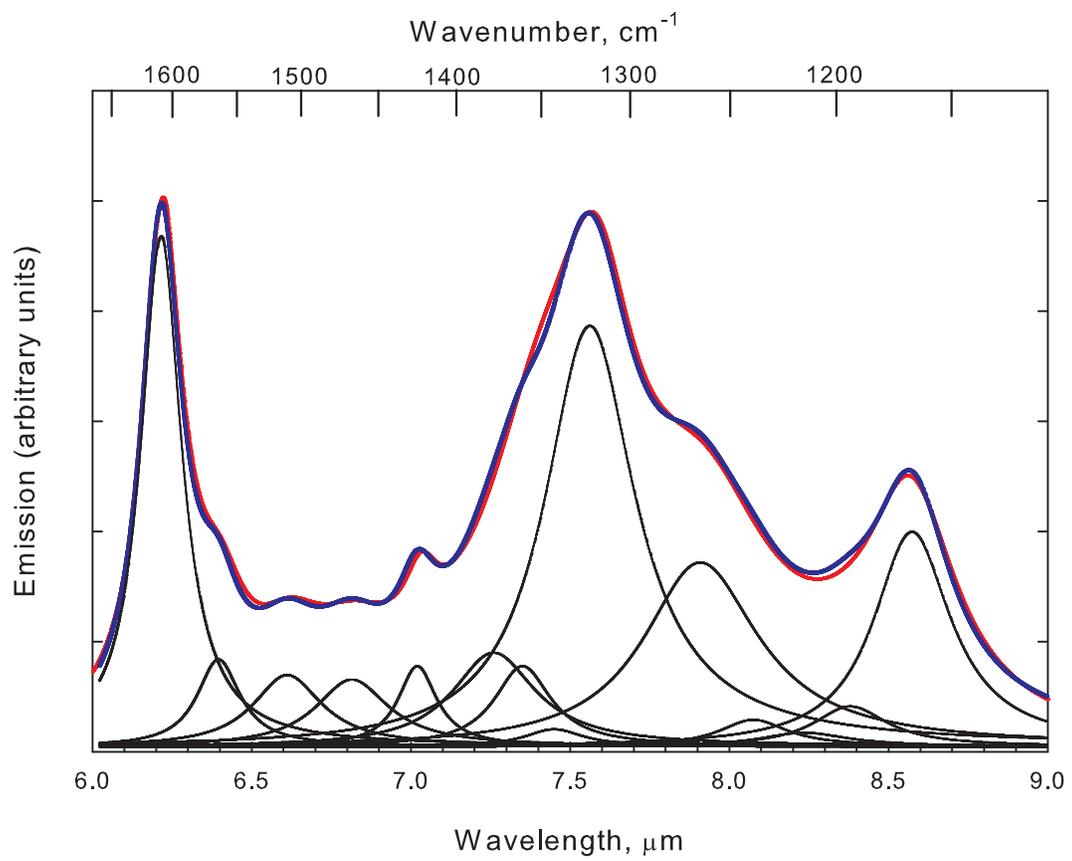

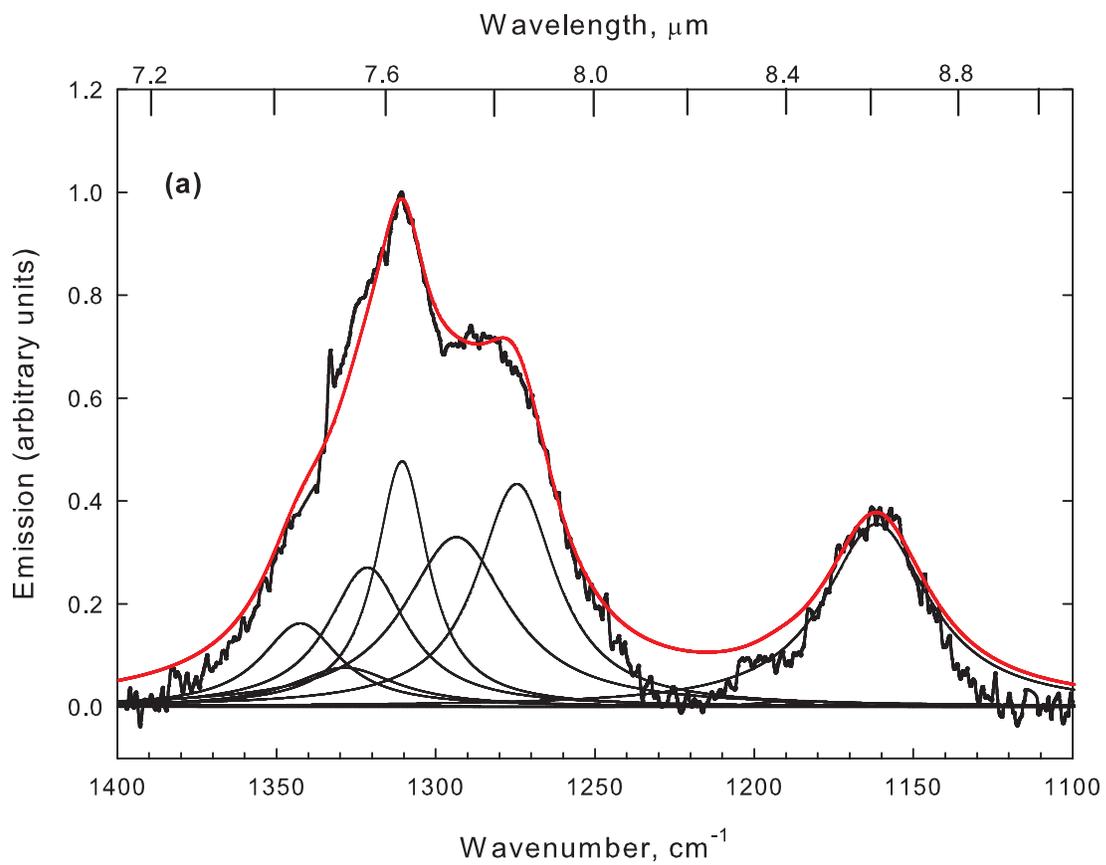

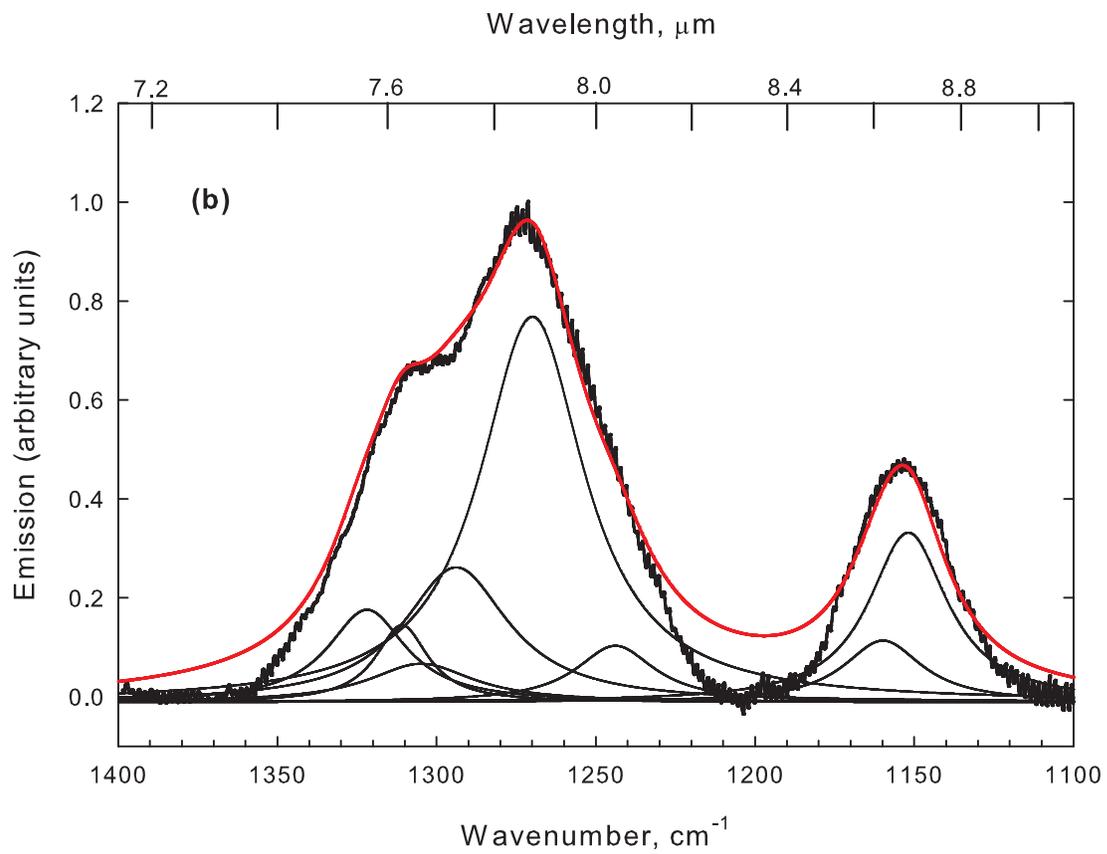

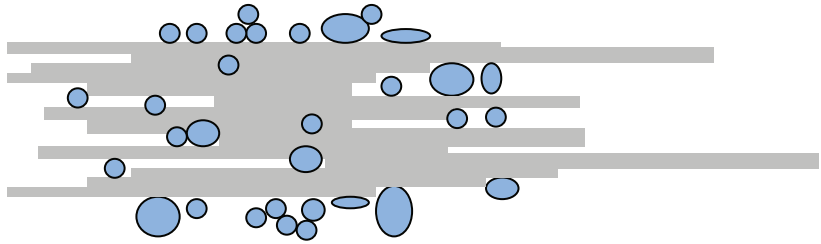